\documentclass{article}
\usepackage{hyperref}
\usepackage{amssymb} 
\usepackage{graphicx}
\usepackage{xcolor}
\usepackage{amsfonts}
\def\ligne#1{\hbox to\hsize{#1}}
\def\leurre{\noindent\leftskip0pt\small\baselineskip 10pt}
\newtheorem{thm}{\textbf{Theorem}}
\newtheorem{fig}{\textbf{Figure}}
\newtheorem{tab}{\textbf{Table}}

\author{Maurice {\sc Margenstern}}
\title{A strongly universal cellular automaton on the heptagrid with six states.}
\begin{document}
\maketitle

\begin{abstract}
In this paper, we prove that there is a strongly universal cellular automaton on the 
heptagrid with six states which is rotation invariant. This improves a previous paper 
of the author with seven states. Here, the structures is slightly simpler and the number of
rules is somehow less.
\end{abstract}

\section{Introduction}~\label{intro}

    In many papers, the author studied the possibility to construct universal cellular
automata in tilings of the hyperbolic plane, a few ones in the hyperbolic $3D$ space. 
Most often, the constructed cellular automaton was weakly universal. By {\it weakly 
universal}, we mean that the automaton is able to simulate a universal device starting
from an infinite initial configuration. However, the initial configuration should not be
arbitrary. It was the case that it was periodic outside a large enough circle, in fact
it was periodic outside such a circle in two different directions as far as the simulated
device was a two-registered machine. In almost all papers, the considered tiling of the
hyperbolic plane was either the pentagrid or the heptagrid, {\it i.e.} the tessellation
$\{5,4\}$, $\{7,3\}$ respectively. Both tessellations live in the hyperbolic plane only.
In the pentagrid, the basic tile is a regular convex pentagon with right angles. In the
heptagrid, it is a regular convex heptagon with the angle 
\hbox{$\displaystyle{{2\pi}\over3}$} between consecutive sides. Below, Figure~\ref{hepta}
provides us with a representation of the heptagrid in the Poincar\'e's disc model of the
hyperbolic plane.

   In the left-hand side picture, we can see seven tiles which are counter-clockwise
numbered from~1 up to~7, those tiles being the neighbours of a tile which we call
the {\bf central tile} for convenience. Indeed, there is no central tile in the heptagrid
as there is no central point in the hyperbolic plane. We can see the disc model as a
window over the hyperbolic plane, as if we were flying over that plane in an abstract
spacecraft. The centre of the circle is the point on which are attention is focused while
the circle itself is our horizon. Accordingly, the central tile is the tile which is 
central with respect to the area under our consideration. The same picture also shows us 
seven arcs of circle. Those arcs represent straight lines in the hyperbolic plane. Those 
lines have the following property: they cross mid-points of consecutive sides of tiles
in the tiling. If two consecutive such mid-points on a line~$\ell$ belong to the same 
tile~$T$, it is not the case for the third mid-point on~$\ell$. That latter point 
belongs to a neighbour~$U$ of~$T$: we say that two tiles of the tiling are 
{\bf neighbours} if and only if they have a common side. In that tiling, if two tiles 
share a vertex they also share a side. That property is entailed by the angle 
\hbox{$\displaystyle{{2\pi}\over3}$}. Lines possessing that latter property are called
{\bf mid-point} lines. Still on that picture, we can see that two mid-point lines
meeting at the mid-point of a side of tile~2 cross mid-points of sides of tile~1. The 
rays which are thus delimited define a sharp angle~$S$, and we call 
{\bf sector headed by~$1$} the set of tiles whose centre are inside~$S$, and tile~1 is 
called the {\bf head} of that sector.
\vskip 10pt
\vtop{
\ligne{\hfill
\includegraphics[scale=1.8]{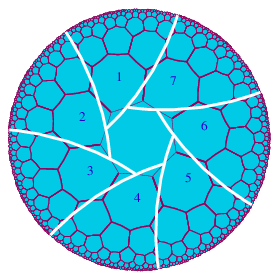}
\raise 13pt\hbox{\includegraphics[scale=0.41]{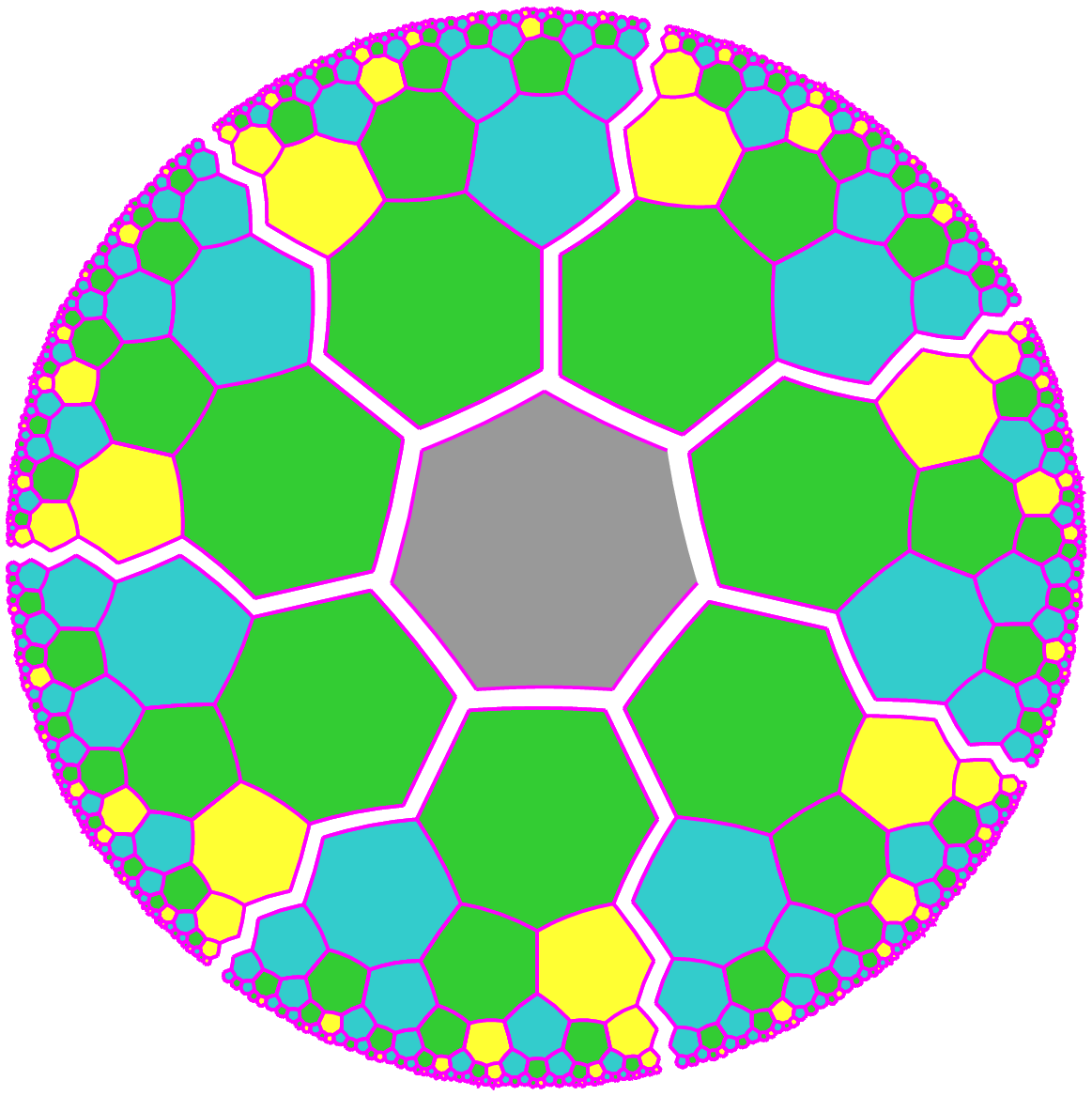}}
\hfill}
\vspace{-15pt}
\begin{fig}\label{hepta}
\leurre
To left: the heptagrid; to right: showing the tree structure of the sectors around a central tile
indicated by the left-hand side picture.
\end{fig}
}

   In the right-hand side picture, we can see the central tile and the seven sectors
which surround it: they are defined by the mid-point lines shown on the left-hand side
picture of Figure~\ref{hepta}. In each sector we have three kind of tiles: green, yellow
and blue ones. If we look from the head of a sector to the tiles close to the horizon,
the head of the sector is green and it has three neighbours in the sector: blue, 
green and yellow while counter-clockwise turning around the head. Blue tiles have two
neighbours which are blue and green in that order and green and yellow tiles have three
neighbours which are blue, green and yellow in that order. These indications define rules
from which we can build a tree which spans the tiles of the sector. More indication
on that topic and on the way to locate the tiles can be found in~\cite{mmbook1,mmJUCS}.

   Now that the global setting is given, we shall proceed as follows: 
Section~\ref{scenario} indicates the main lines of the implementation which is precisely
described in Subsection~\ref{newrailway}. At last, Section~\ref{srules} gives us the rules
followed by the automaton. We refer the reader to \cite{mmarXiv21} for figures which
illustrate the application of the rules. Those figures were established from pieces of 
figures drawn by a computer program which
applied the rules of the automaton to an appropriate window in each of the configurations
described in Subsection~\ref{newrailway}. The computer program also checked that the set
of rules is coherent and that rules are pairwise independent with respect to rotation 
invariance.

   That allowed us to prove the following property:

\begin{thm}\label{letheo}
There is a strongly universal cellular automaton on the heptagrid
which is rotation invariant, truly planar and which has six states. The automaton makes use of
$510$ rules.
\end{thm}

   In paper~\cite{mmarXiv21}, the automaton required 1119 rules and in~\cite{mmarXiv23} with seven 
states too, it required 486 rules.

\section{Main lines of the computation}\label{scenario}

   The first paper about a universal cellular automaton in the pentagrid, the 
tessellation $\{5,4\}$ of the hyperbolic plane, was \cite{fhmmTCS}. This cellular 
automaton was also rotation invariant, at each step of the computation, the set of non 
quiescent states had infinitely many cycles: we shall say that it is a truly planar 
cellular automaton. That automaton had 22~states. That result was improved by a cellular 
automaton with 9~states in~\cite{mmysPPL}. Recently, it was improved with 5~states, 
see~\cite{mmpenta5st}. A bit later, I proved that in the heptagrid, the tessellation 
$\{7,3\}$ of the hyperbolic plane, there is a weakly universal cellular automaton with 
three states which is rotation invariant and which is truly planar, \cite{mmhepta3st}. 
Later, I improved the result down to two states but the rules are no more rotation 
invariant, see~\cite{mmpaper2st}. Paper \cite{JAC2010} constructs three cellular
automata which are strongly universal and rotation invariant: one in the pentagrid, one 
in the heptagrid, one in the tessellation \hbox{$\{5,3,4\}$} of the hyperbolic 
$3D$-space. By strongly universal we mean that the initial configuration is finite, 
{\it i.e.} it lies within a large enough circle.

    In the present paper, we borrow ideas from~\cite{mmarXiv23} and from the previous paper,
\cite{mmarXiv21}.

To~\cite{mmarXiv21}, we borrow the idea of implementing a two register structure which is finite at
each time of the computation. It is constructed as
a segment of straight line which is continued each time the register is incremented
by an appropriate instruction. Here, it is also diminished each time the register is decremented
by an appropriate instruction. A part of the configuration is devoted to the instructions
of the register machine, call it the {\bf program} of the simulation. From the program
to the registers and for the program itself, we borrow the ideas of~\cite{mmarXiv23}.
However, the connection is not that trivial and the technical constraints to remain with
the small number of states entailed many changes in the ideas of 
\cite{mmarXiv21} which allow to strongly reduce the number of rules.

    The simulation is based on the railway model devised in~\cite{stewart} revisited
by the implementations given in the author's paper~\cite{mmarXiv23}.
Sub-section~\ref{railway} describes the main structures of the model. In 
Sub-section~\ref{newrailway} we indicate the new features used in the present simulation.

\subsection{The railway model}\label{railway}

   The railway model of~\cite{stewart} lives in the Euclidean plane. It consists of
{\bf tracks} and {\bf switches} and the configuration of all switches at time~$t$
defines the configuration of the computation at that time. There are three kinds of
switches, illustrated by Figure~\ref{switches}. The changes of the switch configurations
are performed by a locomotive which runs over the circuit defined by the tracks and their
connections organised by the switches.

A switch gathers three tracks $a$, $b$ and~$c$ at a point. In an active crossing,
the locomotive goes from~$a$ to either~$b$ or~$c$. In a passive crossing, it goes
either from~$b$ or~$c$ to~$a$. 

\vskip 10pt
\vtop{
\ligne{\hfill
\includegraphics[scale=0.8]{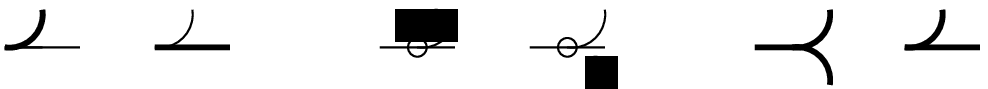}
\hfill}
\begin{fig}\label{switches}
\leurre
The switches used in the railway circuit of the model. To left, the fixed switch, in the
middle, the flip-flop switch, to right the memory switch. In the flip-flop switch, the 
bullet indicates which track has to be taken.
\end{fig}
}

In the fixed switch, the locomotive goes from~$a$ 
to always the same track: either~$b$ or~$c$. The passive crossing of the fixed switch is
possible. The flip-flop switch is always crossed actively only. If the locomotive
is sent from~$a$ to~$b$, $c$ by the switch, it will sent to~$c$, $b$ respectively.
The memory switch can be crossed actively or passively. Now, the track taken by the 
locomotive in an active passage is the track taken by the locomotive in the last
passive crossing.

   As an example, we give here the circuit which stores a one-bit unit of information,
see Figure~\ref{basicelem}. The locomotive may enter the circuit either through the 
gate~$R$ or through the gate~$W$.

  If it enters through the gate~$R$ where a memory switch sits, it goes either through
the track marked with~1 or through the track marked with~0. When it crossed the switch
through track~1, 0, it leaves the unit through the gate~$E_1$, $E_0$ respectively.
Note that on both ways, there are fixed switch sending the locomotive to the appropriate
gate~$E_i$. If the locomotive enters the unit through the gate~$W$, it is sent to the 
gate~$R$, either through track~0 or track~1 from~$W$. Accordingly, the locomotive
arrives to~$R$ where it crosses the switch passively, leaving the unit through the 
gate~$E$ thanks to a fixed switch leading to that latter gate. When the locomotive 
took track~0, 1 from~$W$, the switch after that indicates track~1, 0 respectively and the 
locomotive arrives at~$R$ through track~1, 0 of~$R$. The track are numbered according to 
the value stored in the unit. By definition, the unit is~0, 1 when both tracks from~$W$ 
and from~$R$ are~0, 1 respectively. So that, as seen from that study, the entry 
through~$R$ performs a reading of the unit while the entry through~$W$, changes the unit
from~0 to~1 or from~1 to~0: the entry through~$W$ should be used when it is needed to
change the content of the unit and only in that case. The structure works like a memory
which can be read or rewritten. It is the reason why call it the {\bf one-bit memory}.

\vtop{
\ligne{\hfill
\includegraphics[scale=0.6]{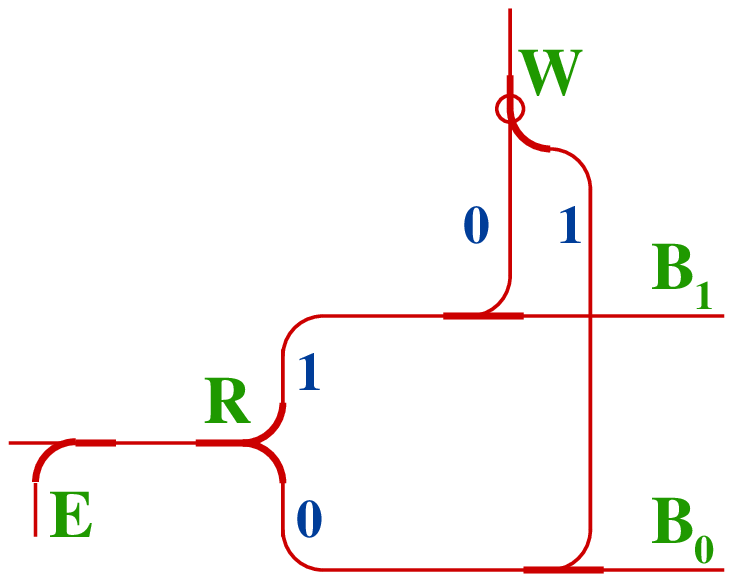}
\hfill}
\begin{fig}\label{basicelem}
\leurre
The basic element containing one bit of information.
\end{fig}
}

   We shall see how to combine one-bit memories in the next sub-section as far as we 
introduce several changes to the original setting for the reasons we indicate there.

\subsection{Tuning the railway model}\label{newrailway}

   We first look at the implementation of the tracks in Sub-subsection~\ref{sbbtracks}
and how it is possible to define the crossing of two tracks.
In Sub-subsection~\ref{sbbswitch} we see how the switches are implemented.
Then, in 
Sub-subsection~\ref{sbbunit}, we see how the one-bit memory is implemented in the new 
context and then, in Sub-section~\ref{sbbregdisc}, how we use it in various places. 

\subsubsection{The tracks}\label{sbbtracks}

    The tracks play a key role in the computation, as important as instructions and 
registers: indeed, they convey information without which any computation is impossible.
   
    As in~\cite{mmarXiv23}, the tracks are one-way. But as in~\cite{fhmmTCS} we define 
the tracks as a particular colour. However, as far as the locomotive goes from the
program to the registers and from there back to the program, it should be possible for
the locomotive to move in opposite directions in some sense. As in~\cite{mmarXiv23}, the
memory switch is split into two different structures: one of them for an active passage, the other
for the passive passage.

Consider four tiles pairwise adjacent, 
the third one being not in contact with the first one. The motion from the first tile to 
the third one can be symbolized as follows:
\def\ftt {\footnotesize\tt}
\vskip 5pt
\ligne{\hfill\ftt FWWW\hskip 30pt RFWW\hskip 30pt WRFW\hskip 30pt WWRF\hskip 30pt WWWR \hfill}

\noindent
and the reverse motion is given by:
\vskip 5pt
\ligne{\hfill\ftt WWWF\hskip 30pt WWFR\hskip 30pt WFRW\hskip 30pt FRWW\hskip 30pt RWWW\hfill}
\vskip 5pt
From that observation, we can deduce that a uniform colour for the tracks can accept a 
single direction, although it can accept both of them as far as the locomotive consists of two
contiguous cells, the front and the rear. The single colour allows us a rather
free use of the tracks. We choose that they follow segments of lines or arcs of circles. Later, we 
display figures illustrating that point.

Consider two tiles~$U$ and~$V$. A {\bf path} from~$U$ to~$V$ is a finite
sequence of tiles \hbox{$\{T_i\}_{i\in[0..n]}$}, where \hbox{$T_0=U$}, \hbox{$T_n=V$}
and, for any $i$ in \hbox{$[1..n$$-$$1]$}, $T_i$ and $T_{i+1}$ share a common side.
In that case, $n$$-$1 is the {\bf length} of the path. The {\bf distance} from~$U$ to~$V$
is the smallest length of the paths from~$U$ to~$V$. A {\bf circle} around~$T$ of
radius~$n$ is the set of tiles~$U$ whose distance to~$T$ is~$n$. An arc of circle is
a path from a tile~$U$ of a circle~$\mathcal C$ to another tile~$V$ of~$\mathcal C$,
whose all tiles also belong to~$\mathcal C$. 
\def\BB{{\tt B}}
\def\GG{{\tt G}}
\def\RR{{\tt R}}
\def\YY{{\tt Y}}
\def\WW{{\tt W}}
The colour of the tracks is fixed as \YY. However we shall indicate a track with respect to the 
colour of the locomotive which runs on that track. A \BB-, \GG-track is run by a locomotive whose
front is \BB-, \GG- respectively. We also say \BB-, \GG-locomotives respectively. Paths from a part
of the circuit to another part consists of pieces of tracks connected together. On most parts of the
circuit, the locomotive is \BB-. Later, we shall say that \BB- and \GG- are {\bf opposite colours}.

\vskip 10pt
\vtop{
\ligne{\hfill
\includegraphics[scale=1.7]{secteurs.ps}
\includegraphics[scale=1.7]{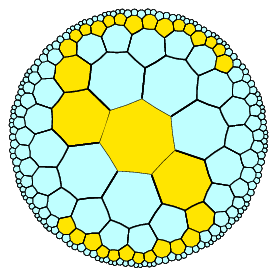}
\hfill}
\begin{fig}\label{f_voies}
\leurre
To right, the tracks used for testing the implementation. To left, the sectors in order to locate
the tiles of the tracks.
\end{fig}
}
\vskip 10pt
The right-hand side part of Figure~\ref{f_voies} represents a track used to test the implementation
of the tracks. The left-hand side of the figure reproduces the sectors illustrated in
Figure~\ref{hepta} in order to locate the tiles of the tracks. From top to bottom, those tiles
are (7,7) up to (7,12), then (1,5) to (1,12), then (2,2), (2,1), (0,0), (5,1), (5,3), (5,7), (5,6),
(5,5), and then (4,12) down to (4,5). The locomotives, both \BB- and \GG- can run on that track
from (7,7) down to (4,5) and also in the reverse order, from (4,5) up to (7,7): it has been tested
by a computer program.

   We now turn to Sub-subsection~\ref{sssauxil} where we describe the auxiliary structures
and the passive fixed switch used to implement the crossings and then the remaining switches whose 
implementation is discussed in Sub-subsection~\ref{sbbswitch}.

\subsubsection{Auxiliary structures and crossings}\label{sssauxil}

The auxiliary structures we need are the fork, the changer and the filter. The changer and the 
filter are implemented in two versions : a \BB- and a \GG-one. The \BB, \GG-changer transforms a
\GG-, \BB-locomotive respectively into a \BB-, \GG-one respectively.  The \BB-, \GG-filter
let a \BB-, \GG- locomotive respectively run further and it destroys a \GG-, \BB-locomotive
respectively.

\vskip 10pt
\vtop{
\ligne{\hfill
\includegraphics[scale=1]{secteurs.ps}
\includegraphics[scale=1]{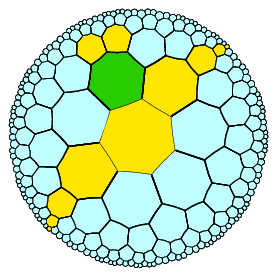}
\includegraphics[scale=1]{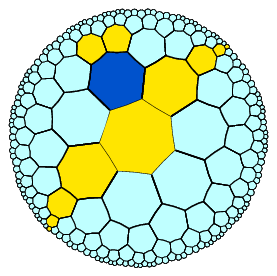}
\hfill}
\ligne{\hfill
\includegraphics[scale=1]{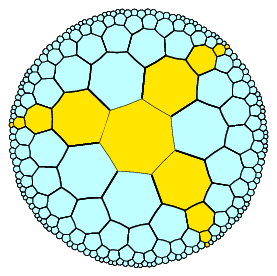}
\includegraphics[scale=1]{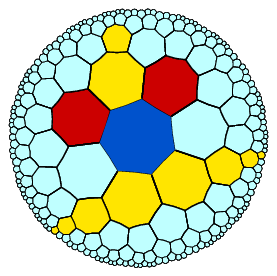}
\includegraphics[scale=1]{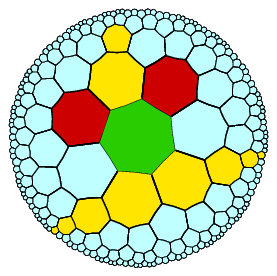}
\hfill}
\begin{fig}\label{f_auxil}
\leurre
Top, from left to right: the sectors, the changer from \BB- to \GG-, the changer from \GG- to \BB.
Bottom, from left to right: the fork, the filter for \BB-, the filter for \GG-.
\end{fig}
}
\vskip 10pt

The figure illustrates the {\bf idle configuration} of those structure. We define an idle 
configuration of a structure to be a structure contained in a disc of radius~2 around the central
tile in which disc there is no locomotive.

In the changers, a track crosses the figure through tiles (7,21), (7,8), (7,3), (7,1), 0, (3,1),
(3,3), (3,8) and (3,21). In (1,1) we have a \BB- or a \GG-tile and in (1,3) and (1,4) we have a
\YY-tile which fixes the tile in (1,1). In the filters,the track crosses the figure through the 
tiles (6,20), (6,7), (6,2), (5,1), (4,1), (4,2), (4,5) and (4,13). Tile~0 is \BB- or \GG-. It
has \RR-tiles at (2,1) and (7,1) and there are two \YY-tiles at (1,1) and (1,3) whose role is
depicted later.

At last, the fork gathers three tracks around a central \YY-tile: the arriving track passing through
(5,21), (5,8), (5,3) and (5,1), the track leaving to left through (2,1), (2,4), (2,12) and (2,33)
and the track leaving to right through (7,1), (7,3), (7,8) and (7,21).

A computer program checked the motion of the locomotive through each of those structure and checked
that the structures worked as far as it was expected from them to do.
\vskip 10pt
The fixed switch is crossed passively only in the present implementation, 
following~\cite{mmarXiv23}. 

\vskip 10pt
\vtop{
\ligne{\hfill
\includegraphics[scale=1.7]{secteurs.ps}
\includegraphics[scale=1.7]{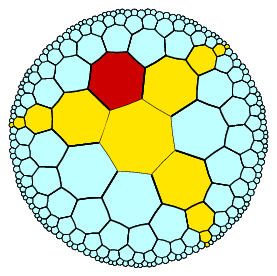}
\hfill}
\begin{fig}\label{f_i_fix}
\leurre
\end{fig}
}
\vskip 10pt
Note the occurrence of an \RR-tile in (1,1). It allows the passage of a locomotive along one of its
sides while it prevents the front of the locomotive to enter the branch starting from its side 
where the other branch of the switch arrives.
\vskip 10pt
The fork, the changers, the filters and the fixed switch can be assembled in a structure which 
performs the working of a crossing. That organisation is illustrated by Figure~\ref{f_croise}.
\vskip 10pt
\vtop{
\ligne{\hfill
\includegraphics[scale=0.5]{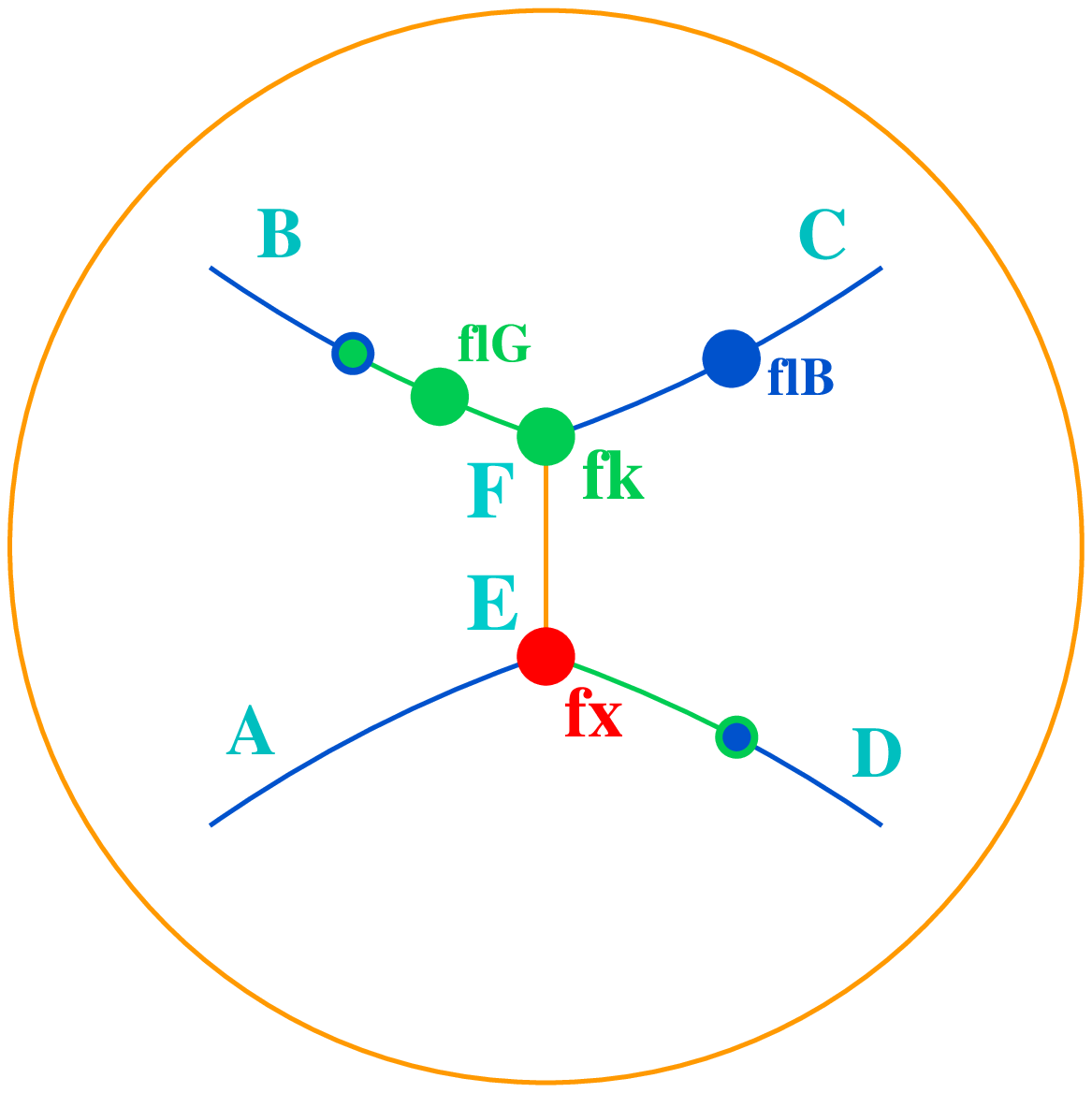}
\hfill}
\vspace{-20pt}
\begin{fig}\label{f_croise}
\leurre
Organisation of a crossing.
\end{fig}
}

A locomotive arrives at~$C$. If it comes from~$A$, $D$ it goes to~$C$, $B$ respectively. If the
locomotive has a given colour, it is assumed that it keeps its colour along~$C$ where, after~$E$
there is a filter which let those locomotives pass. If the locomotive arrives from~$D$, before 
reaching~$E$, a changer transforms that locomotive in a new locomotive with the opposite colour.
That new locomotive is supposed to go to~$D$ where an appropriate filter is sitting.  At $E$ we
have a passive fixed switch which allows the locomotive, whatever its colour is, to reach the
fork sitting at~$E$. That fork sends a copy of that locomotive, with the same colour, on each
branch, to~$B$ and to~$C$. Now, the filter let only pass the locomotive whose colour is that of the
filter. As far as the locomotive coming from~$D$ has changed its colour before reaching~$E$, the
opposite changer will allow that locomotive to recover the initial colour it as on the $D$ track.

\subsubsection{The switches}\label{sbbswitch}

   The present sub-subsection deals with the flip-flop switch and with the memory one.

   In an active crossing of a switch where the locomotive arrives through $a$, we say that $b$, 
$c$ is the {\bf selected} track if the locomotive goes from~$a$ to~$b$, to~$c$ respectively. A 
flip-flop changes the selection after an active crossing and there is never a passive crossing of a
flip-flop switch. In a memory switch, its passive part defines the selection of
its active part. Accordingly, we need two kinds of structures: a structure which let
the locomotive go, the other one which kills it. In the flip-flop, the change is
made by the crossing of the switch, in the memory switch the change is triggered by
the crossing of the passive switch. Basically, it is the same constraint. 
Accordingly, the controlling structure in the flip-flop and in the memory switch should be 
{\it programmable}, a point which we explain further.

Figure~\ref{activswitches} illustrates the implementation of the flip-flop, left-hand side
of the figure, ans also the active memory switch, right-hand side of the figure.
Figure~\ref{memoswitches} reminds us the active memory switch and shows us the
passive memory switch on its right-hand side part.

In the flip-flop, the simple locomotive arrives to~$C$. There, a {\bf fork} sends two 
simple locomotives: one towards~$L$, the other towards~$R$. The locomotive which is sent to~$R$ 
where a filter is sitting. If the filter has the colour of the locomotive, that latter goes on.
Otherwise, the locomotive is destroyed. Now, if the filter at~$R$ is \BB-, \GG-, the filter at~$L$
is \GG-, \BB- respectively. Accordingly, the locomotive goes on further on one side of the
switch only, which defines the selected track: it is the leaving track on which the filter has
the colour of the locomotive arriving at~$C$. When the locomotive goes on along the selected track,
at some distance of the filter, it meets a fork: one branch of the fork let the locomotive go on
along the selected track, the other sends the locomotive to~$S$. There, a fixed switch is sitting
and any locomotive arriving there passively crosses the switch. From~$S$, the locomotive arrives
to a fork which sends a copy of it to the filter at~$L$ and another copy to the filter at~$R$. 
Those copies change the filter to the opposite colour, so that now, the selected track of the switch
is changed which is conformal to the definition of a flip-flop switch. Such a change will be
explained in Section~\ref{srules} devoted to the rules observed by the cellular automaton.

Note that the track from $L$ to $R$ is a segment of line while the track from the fork placed after
the filter to~$S$ is an arc of a circle. The change of the filters must occur  after the passage of 
the locomotive through the filter of the selected track. It is enough to define the circle 
supporting that arc with a radius equals to the distance from~$C$ to~$L$ and so that the angle of
the arc should be at least $\displaystyle{\pi\over 2}$: the perimeter of a circle is an exponential
of its radius.

\vskip 10pt
\vtop{
\ligne{\hfill
\includegraphics[scale=0.35]{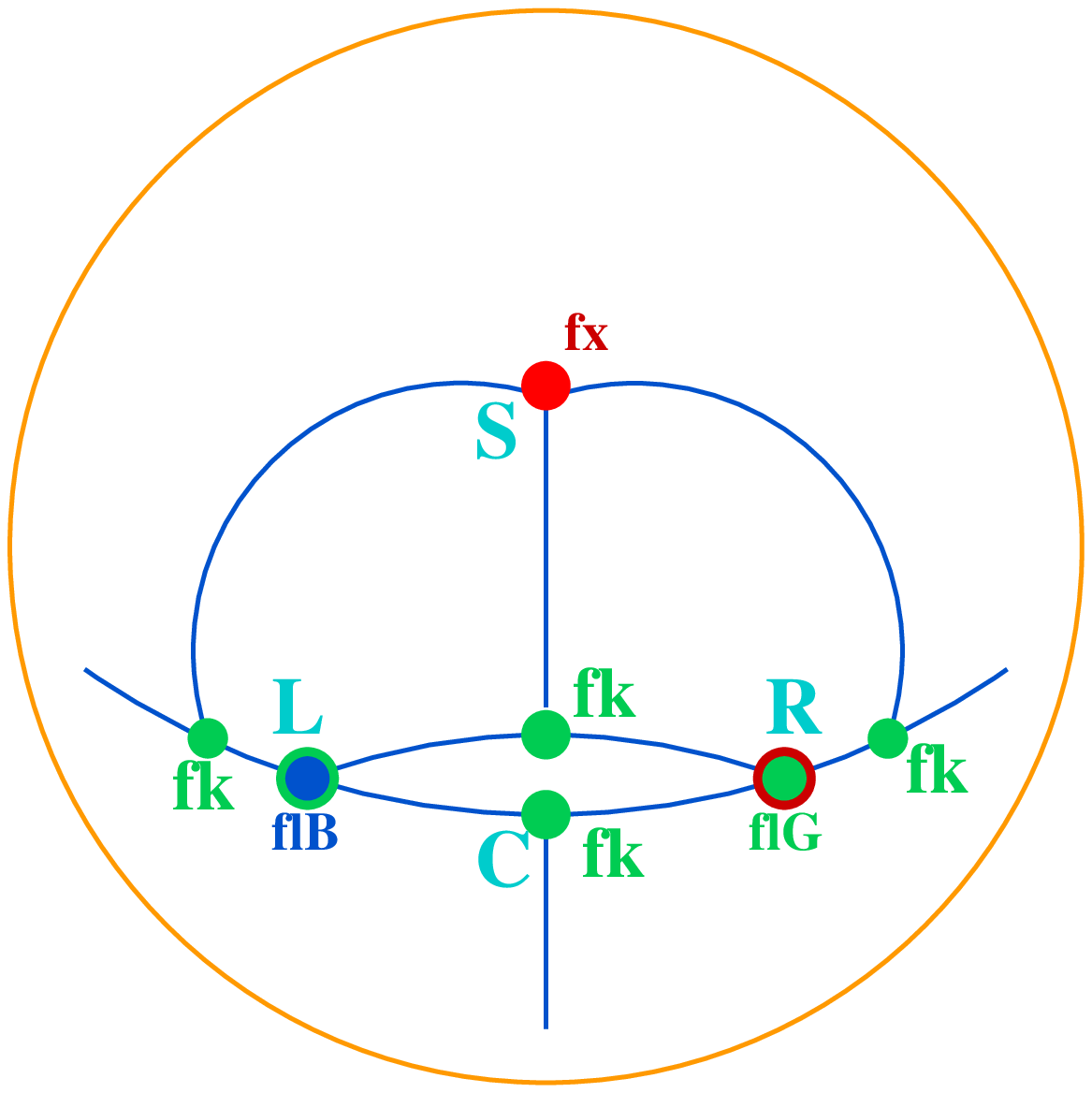}
\includegraphics[scale=0.35]{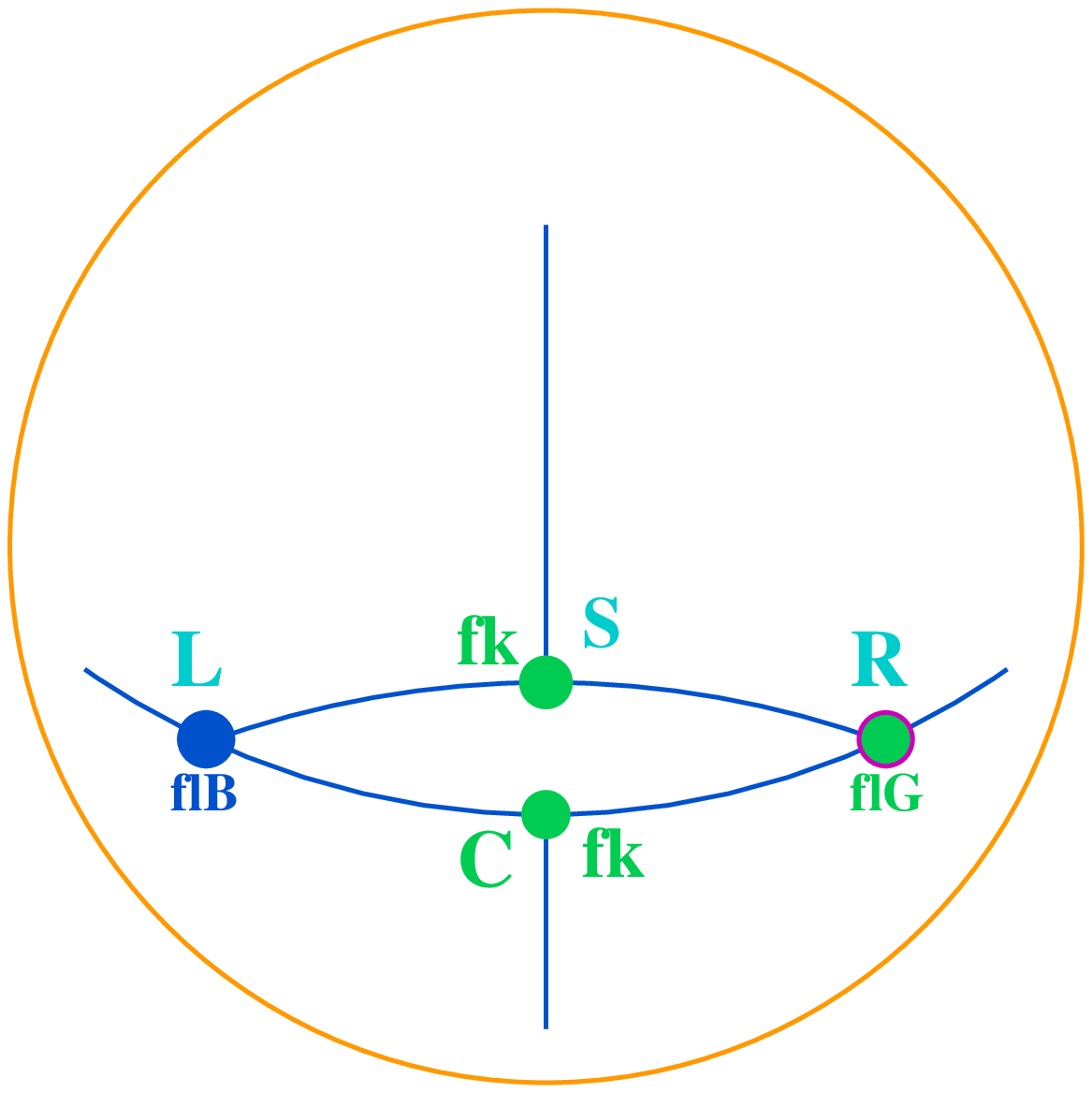}
\hfill}
\begin{fig}\label{activswitches}
\leurre
To left: the flip-flop. To right: the active memory switch.
\end{fig}
}
\vskip 10pt
   The active memory switch is simpler as far as in its working when the locomotive 
arrives at~$C$ only the arcs from~$C$ to~$L$ and from~$C$ to~$R$ are concerned. We may 
assume that a passive crossing occurs when the locomotive is not involved in the
active switch.

\vskip 10pt
\vtop{
\ligne{\hfill
\includegraphics[scale=0.35]{disque_memoact.ps}
\includegraphics[scale=0.35]{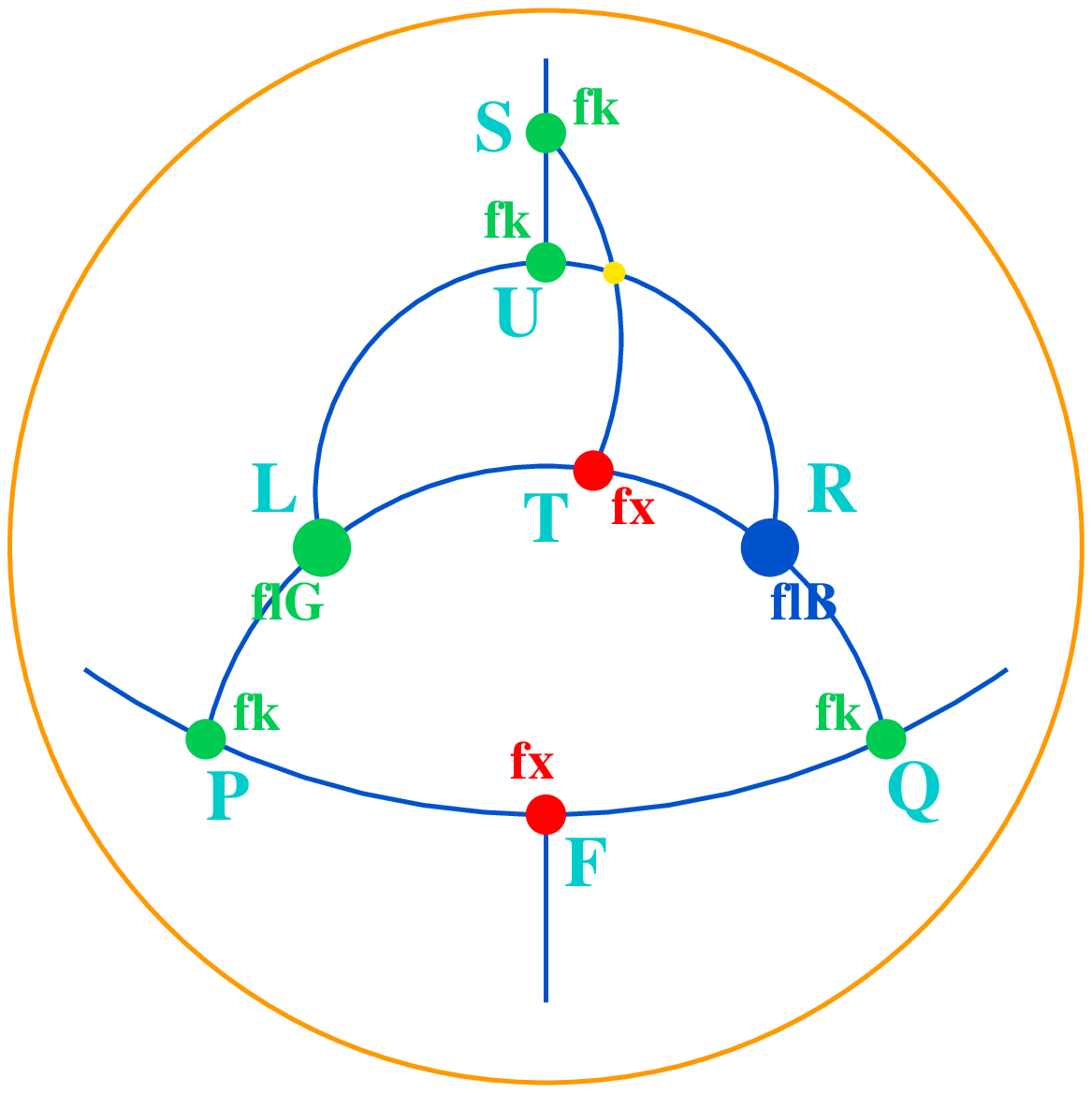}
\hfill}
\begin{fig}\label{memoswitches}
\leurre
Memory switch: to left, the active switch; to right the passive one.
\end{fig}
}

   The structure of the passive switch is very different. The simple locomotive arrives 
either at~$P$ or at~$Q$. Assume that it arrives at~$P$. The case of an arrival at~$Q$ 
is symmetrically dealt with. A fork at~$P$ sends a locomotive to~$F$, a fixed switch which 
let the locomotive leave the switch. The other locomotive is sent to~$L$. There, if the 
filter in the active part let the locomotive go, the filter at~$L$ stops the 
locomotive, so that the memory switch does not change the selection. If the filter
of the active part stops the locomotive, the filter of the passive switch let the 
locomotive go. The locomotive goes further to~$S$ via $T$. There, at~$S$, a fixed switch sends it 
to~$U$ where a fork sends a copy of the locomotive to the fork~$S$ of the active switch
and according to the scheme of the active switch that action will exchange the role of
its filters. The second locomotive sent by the fork at~$S$ goes to~$U$ where another
fork duplicates that locomotive sending one copy to~$L$ and the other to~$R$, both
copies exchanging the roles of the corresponding filters. Accordingly, both parts
of the memory switch operates on the needed way. Here too, we have to make sure that
the arrival at~$L$ and~$R$ of the locomotives sent from~$T$ occurs later at the
permissive controller than the locomotive arriving from~$P$ or from~$Q$. The picture
of Figure~\ref{memoswitches} ensures us that it is easy to fulfill that condition, using arc of a 
circle with an enough large radius.

    Note that the separation of the two parts of the memory switch allows to use one passive part
for changing several active parts, a situation we shall soon encounter.

\subsubsection{The one-bit memory}\label{sbbunit}

   We are now ready to implement the one-bit memory in our setting.
The implementation is illustrated by Figure~\ref{fhypunit} which reminds us the
Euclidean implementation in its left-hand side part.
We adopt the same notations as in the Euclidean picture in order to facilitate the
comparison for the reader.
\vskip 5pt
\vtop{
\ligne{\hfill
\raise 40pt\hbox{\includegraphics[scale=0.6]{elem_gb.ps}}
\hskip-10pt
\includegraphics[scale=0.45]{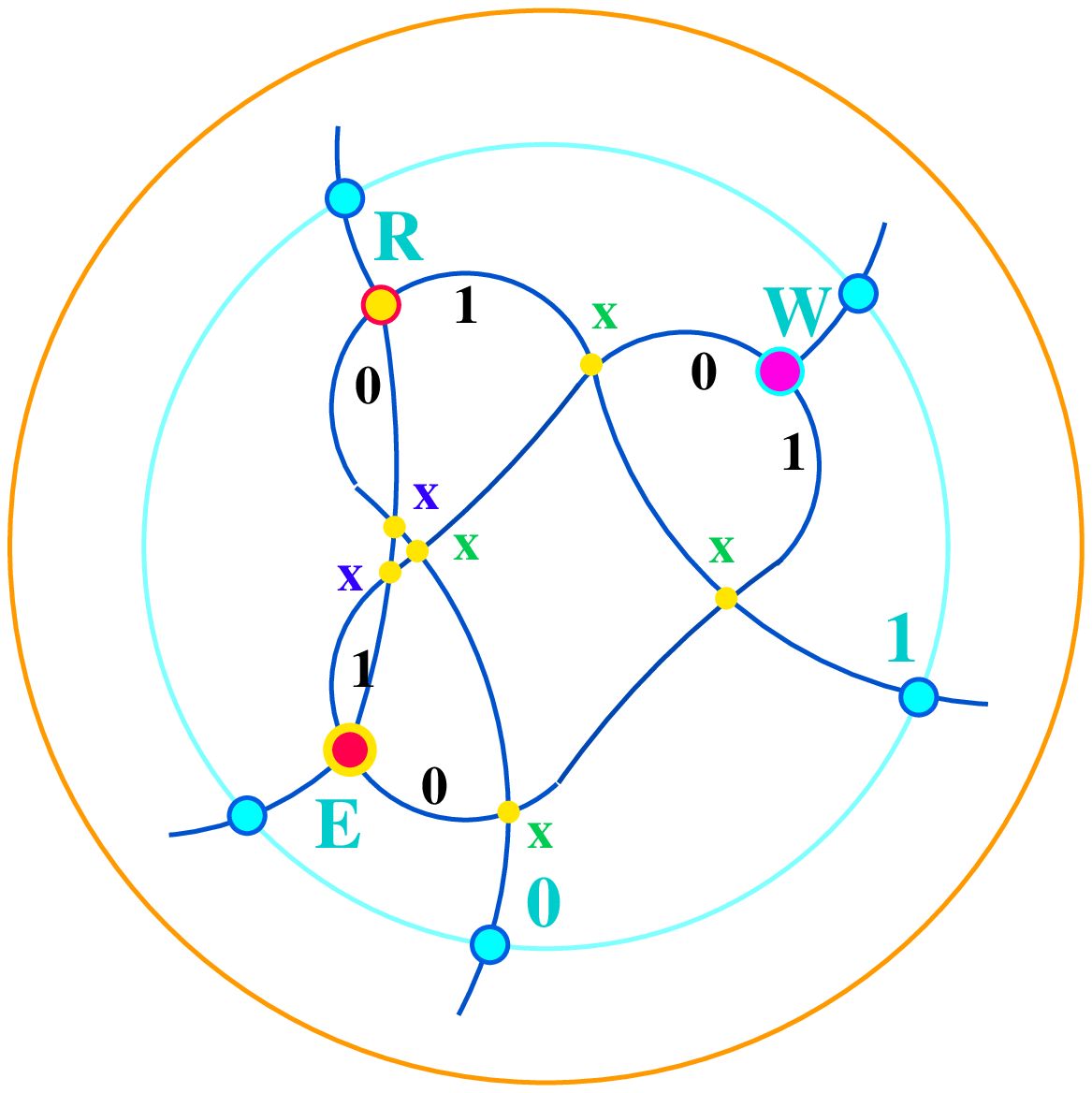}
\hfill}
\vspace{-30pt}
\begin{fig}\label{fhypunit}
\leurre
To left, the scheme of Figure~{\rm{\ref{basicelem}}}, to right, its implementation
in our setting. In the hyperbolic picture, the colours at~$W$, $R$ and~$E$, namely
yellow, light green and light purple respectively represent a flip-flop switch, an
active memory switch and a passive memory switch. The yellow points with an {\tt x}
letter indicate crossings.
\end{fig}
}

Here again, a simple locomotive is involved. If it arrives at the blue disc labelled with~$R$ it 
goes to the red~disc labelled with the same letter and then it exits through the gate~$B0$ or 
through the gate~$B1$ according to the track selected by the active memory switch. 

If the locomotive arrives at the blue~$W$, it is sent to the red~$W$ where a flip-flop 
switch sends the locomotive to the red~$E$ either through track~0 or track~1, depending 
on what is selected by the switch at the red~$R$. If the track is~$i$, then it arrives 
to the red~$E$ through its track $1$$-$$i$. Now, from what we have seen on
Figure~\ref{memoswitches}, the red~$E$ sends a locomotive to the red~$R$ making the
switch select the values which are compatible with those at the red~$E$. But from the
red~$E$, another simple locomotive is sent out of the switch through the blue gate~$E$.
Note that the memory switch of the left-hand side picture is split into two parts on the 
right-hand side figure. There, the active part sits at~$R$ while the passive one sits at~$E$,
a path joining~$E$ to~$R$.

\def\DDI{\hbox{$\mathbb {DD}$$_{\mathbb I}$}}
\def\DDD{\hbox{$\mathbb {DD}$$_{\mathbb D}$}}
\def\DDT{\hbox{$\mathbb {DD}$$_{\mathbb T}$}}
\def\EE{{\bf\tt E}}
\def\zz{{\bf\tt 0}}
\def\uu{{\bf\tt 1}}
As far as we shall later several times use the one-bit memory, we shall represent it by a blue disc 
with five gates labelled with \RR, \EE, \WW, \zz{} and \uu.

\subsubsection{A register and its discriminating structures}\label{sbbregdisc}

\def\MM{{\tt M}}
Figure~\ref{f_reg} illustrates the implementation of a register in our setting.
As can be seen on the figure, we have a segment of line enclosed in a path surrounding it. 

   The value of the register is the number of \WW-tiles from tile~0{} in the left-hand side of
Figure~\ref{f_reg} until tile~0 of the right-hand side part, both tiles included. An
empty register, whose value is~0 by definition, is characterised by the fact that tile~0 at the
beginning is \MM-.

\vskip 10pt
\vtop{
\ligne{\hfill
\includegraphics[scale=1.7]{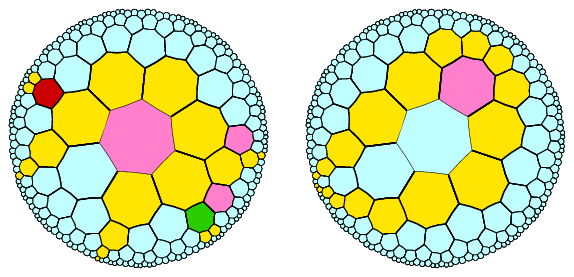}
\hfill}
\vspace{-5pt}
\begin{fig}\label{f_reg}
\leurre
To left, the beginning of a register when $n=0$; to right, its end when $n>0$. Both ends are 
marked as far as they play an important role.
\end{fig}
}

   The register is the occasion to speak of the fact we have two kinds of locomotives. In fact,
the \BB-locomotive is devoted to implement the register while the \GG-one is devoted to
decrement the register. However, the determination of the colour of the locomotive is fixed
by a n appropriate structure before entering the register. Until that point is reached, the
locomotive is \BB-, whatever its role, whether it conveys an instruction or whether it is a
signal to change the working of a filter so that at certain times, three locomotives are 
running in the circuit. Also, we need three new structures outside a register: a structure to
remember which instruction required to increment the register, the \DDI-structure, 
see~\ref{memo_incr}.a; a structure to remember which instruction required to decrement the register,
the \DDD-structure, see~\ref{memo_decr}.b; a structure to remember the type of the instruction,
the \DDT-structure, see \ref{memo_type}.c\ {\it i.e.} to increment or to 
decrement the register. Those structures are required for each 
register. All those structures make use of one-bit memories.

\label{memo_incr}
{\bf \ref{memo_incr}.a\ Remembering the incrementing instruction}

   As can be guessed from Figure~\ref{f_reg}, when the locomotive goes back from the register,
as far as we have two types of locomotive only, the locomotive does not remember from which point
of the program it was issued. In order to go back to that point in order to define the next 
instruction, the instruction must be marked in some way near the register itself. The \DDI-structure
is devoted to play that role. There a \DDI-structure for each register $\mathcal R$. The structure
contains as many units as there are instructions in the program to specifically 
decrement $\mathcal R$. From each instruction $I$ incrementing $\mathcal R$, a path goes from the 
place of $I$ in the program to the unit of the \DDI{} attached to $\mathcal R$ associated by that
path to that unit. The unit itself contains a one-bit memory, see Figure~\ref{f_uDDI}.

\vtop{
\ligne{\hfill
\includegraphics[scale=0.5]{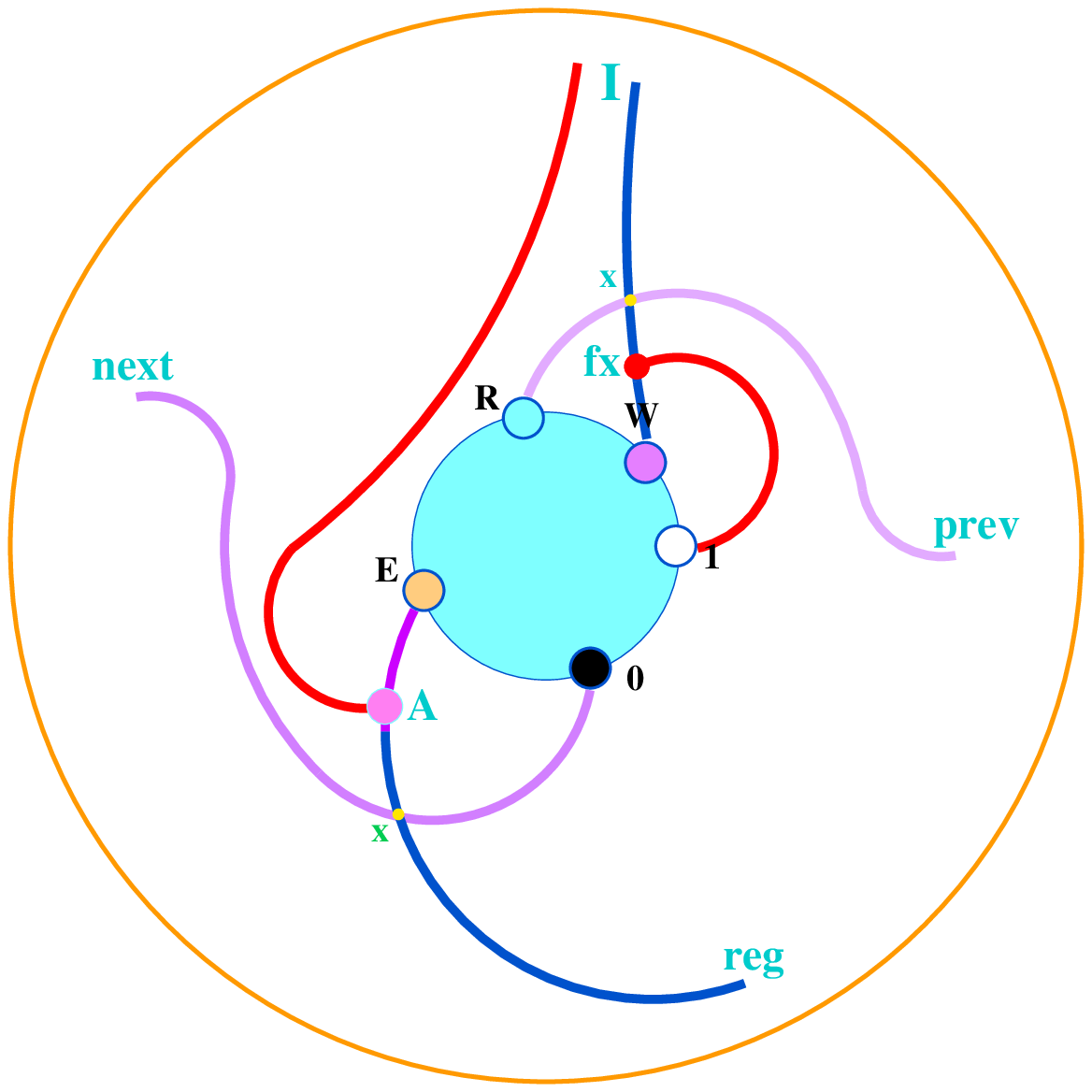}
\hfill}
\begin{fig}\label{f_uDDI}
\leurre
The unit of \DDI{} devoted to an incrementing instruction.
\end{fig}
}

At the initial configuration, in all units of each \DDI-structure, the one-bit memory contains~\zz.
When the locomotive starts its way to perform an incrementing instruction~$I$ over the register
$\mathcal R$, the track run by the locomotive arrives to the \WW-gate of the unit of the \DDI{}
attached to $\mathcal R$ which is associated with~$I$. As far as the locomotive arrives to a 
\WW-gate it rewrites the \zz-bit of the one-bit memory to \uu. Then, the locomotive exit the memory
through its \EE-gate. It meets a flip-flop at~$A$ which, form the initial configuration, sends it
towards the \DDT-structure of $\mathcal R$. After crossing that flip-flop switch, the selected
track is now a track which goes back to the program, to the instruction which has to be
executed after $I$ has completed its operation on $\mathcal R$. That track is in orange on
Figure~\ref{f_uDDI}. At that time, there is a single unit in that \DDI{} whose one-bit memory
contains \uu.

When the locomotive comes back from $\mathcal R$, it visits the \DDT-structure which remembers that
it performed an incrementing instruction so that it sends the locomotive to the \DDI{} of
$\mathcal R$.

The locomotive visits each unit of the \DDI-structure until it meets the single one whose one-bit 
memory contains~\uu. To do that, the locomotive enters the memory of the unit through its \RR-gate.
If it reads~\zz, it is sent to the next unit. When it reads~\uu, it knows that the right unit is
reached. Leaving the one-bit memory through the \uu-gate, the locomotive is sent to the \WW-gate so
that it rewrites the content of the one-bit memory from~\uu{} to \zz. Leaving the memory through the\EE-gate, the locomotive again meets $A$ where the selected track of the flip-flop sends it on the
track leading to the right place of the program. Once the flip-flop at $A$ is crossed, the selected
track again becomes the track leading to $\mathcal R$. Accordingly, when the locomotive leaves the 
\DDI-structure, that one recovered its initial configuration.

\label{memo_decr}
{\bf \ref{memo_decr}.b\ Remembering the decrementing instruction}

    If the locomotive has to decrement the register $\mathcal R$ it visits a similar structure
as \DDI, the \DDD-structure attache to $\mathcal R$. The principle of that structure is similar to 
that of \DDI. However a unit of \DDD{} is more complex than a unit of \DDI. The reason comes from
the difference between a decrementing instruction and an incrementing one. It is always possible to
increment a register. It is not possible to decrement an empty register as far as register machines
deal with natural numbers only. When a decrementing locomotive arrives at the beginning of
an empty register $\mathcal R$, it leaves the register through a different track from which it 
leaves the register having successfully decremented it.

\vtop{
\ligne{\hfill
\includegraphics[scale=0.5]{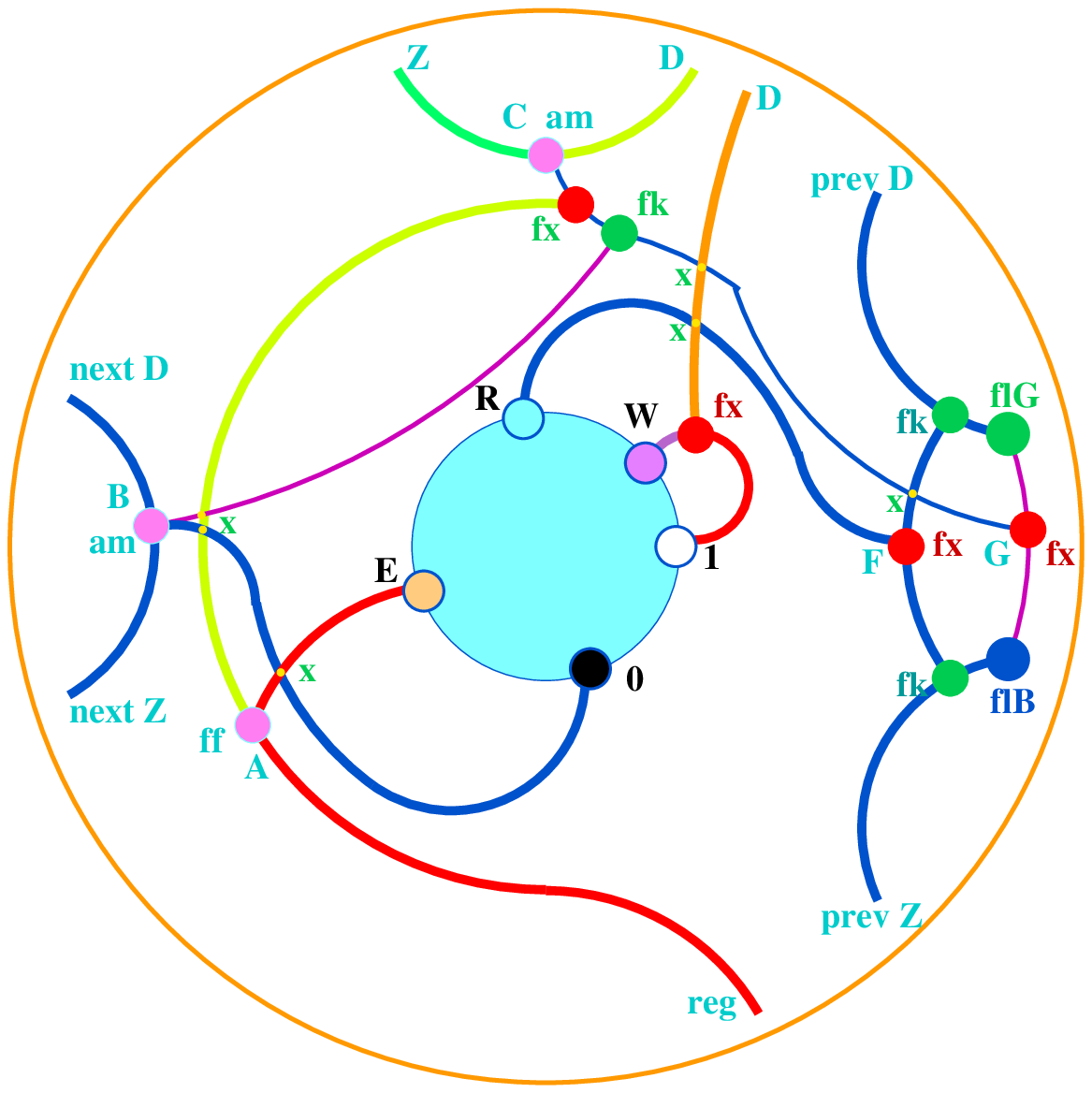}
\hfill}
\vspace{-30pt}
\begin{fig}\label{f_uDDD}
\leurre
The unit of \DDD{} devoted to a decrementing instruction.
\end{fig}
}

   Accordingly, if the working of \DDD{} is similar to that of \DDI{} for an arriving locomotive 
from the program, it is not the case for a locomotive returning from the register. We leave the
case of an arriving locomotive from the program to the reader.

\def\ZZZ{{\bf\tt Z}}
\def\DD{{\bf\tt D}}
   Consider the case of a locomotive coming back from~$\mathcal R$. It comes from the \DD-track
after a successful operation or it comes through the \ZZZ-track because it could not decrement an 
empty $\mathcal R$. Both tracks behave in the same: they lead to the \RR-gate. If the locomotive
reads \zz, it goes to the next unit through the same kind of track as the one it used to arrive at 
the current unit. As can be seen, there are two possible exits for the locomotive entering the
one-bit from through the \RR-gate when it comes from the register. But in both cases, the exit
splits again in two possibilities depending on which track the locomotive entered the unit.
In case of an exit to the next unit, the exit track is a \DD-, \ZZZ-track if it was respectively 
the case for the entering track. In case of an exit to the program, the exit rack goes to the next
instruction or to another one defined by the jump condition it the entering track was \DD- or \ZZZ-
respectively. The distinction is known at the entry of the unit, see point $G$ in 
Figure~\ref{f_uDDD}. It is the reason why at $G$ a passive memory switch is sitting. As far as
that difference has to be known at~$B$ and at~$C$, there is at each of those points an active
memory switch connected to the passive one at~$G$. The connection to those active memory switches
is established by a track from~$G$ to a fork before reaching~$C$ which sends a copy to~$C$ and 
another copy to~$B$. Accordingly, the whole structure performed what is expected from it in all 
situations.

\label{memo_type}
{\bf \ref{memo_type}.c\ Remembering the type of instruction}

The working of that structure reminds us that of \DDD. Here two, there are two possible tracks 
arriving to the structure: either from a \DDI-structure or from a \DDD-one.

\vtop{
\ligne{\hfill
\includegraphics[scale=0.5]{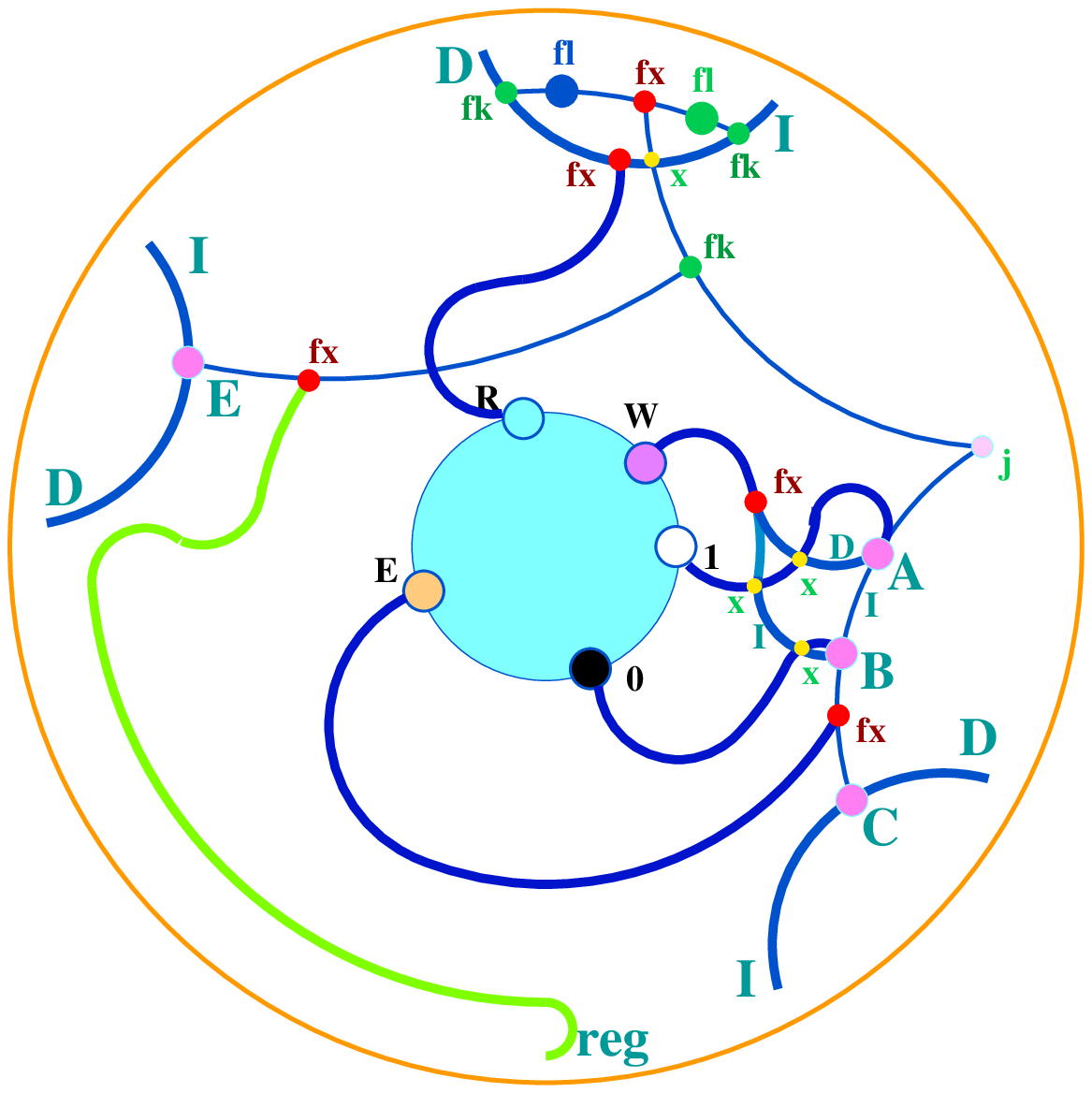}
\hfill}
\vspace{-30pt}
\begin{fig}\label{f_uDDT}
\leurre
The unit of \DDT{} devoted to recover the type of the instruction.
\end{fig}
}

   There are also two possible exits: one back to the \DDI- or \DDD-structure, the other to the
register $\mathcal R$. But in that latter case,there are two different tracks depending on the 
instruction: whether it has to increment $\mathcal R$ or to decrement it. The discrimination is
fixed by a passive memory switch on the arrival point to the structure: it is illustrated in
the upper part of Figure~\ref{f_uDDT}. That piece of information is transferred to the exit part
of \DDT: to the register and to \DDI{} or \DDD. On figure~\ref{f_uDDT}, we can see a track from the 
arrival to~$C$ via~$j$ and to~$E$ thanks to a fork which is met by the locomotive before reaching
$j$. At $C$ and at~$E$ an active memory switch is sitting connected with the signal possibly
emitted by one of the filters close to the arrival tracks to \DDT. The information about the type
of instruction is also dispatched to other active memory switches at~$A$ and at~$B$. The reason
of those additional memory switches is that reading~\zz{} or~\uu{} has not the same meaning 
according to which instruction is involved. Let us fix that \uu, \zz{} is attached to an instruction
which has to increment, to decrement respectively the register.

\vtop{
\ligne{\hfill
\includegraphics[scale=0.4]{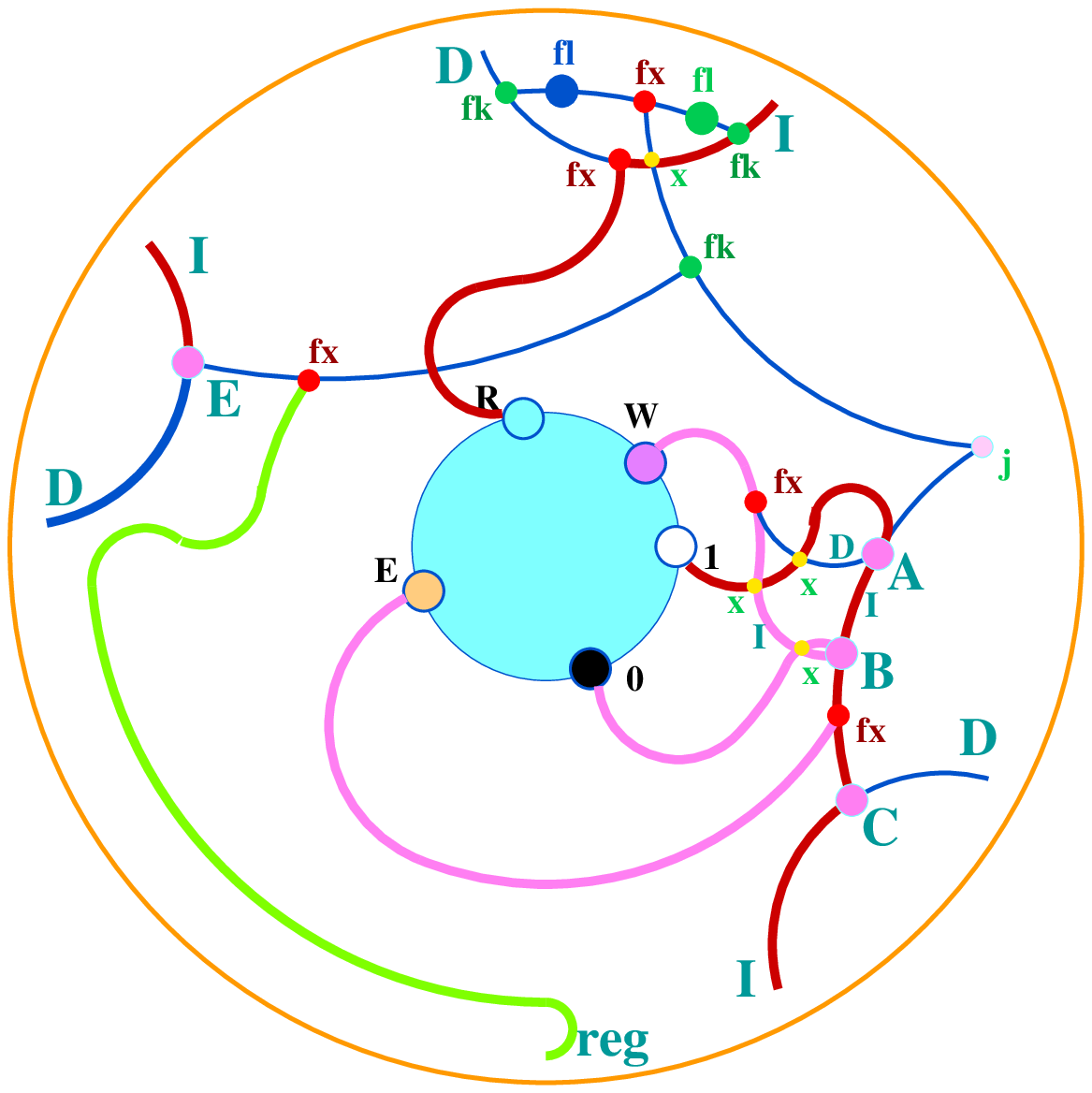}
\hskip-10pt
\includegraphics[scale=0.4]{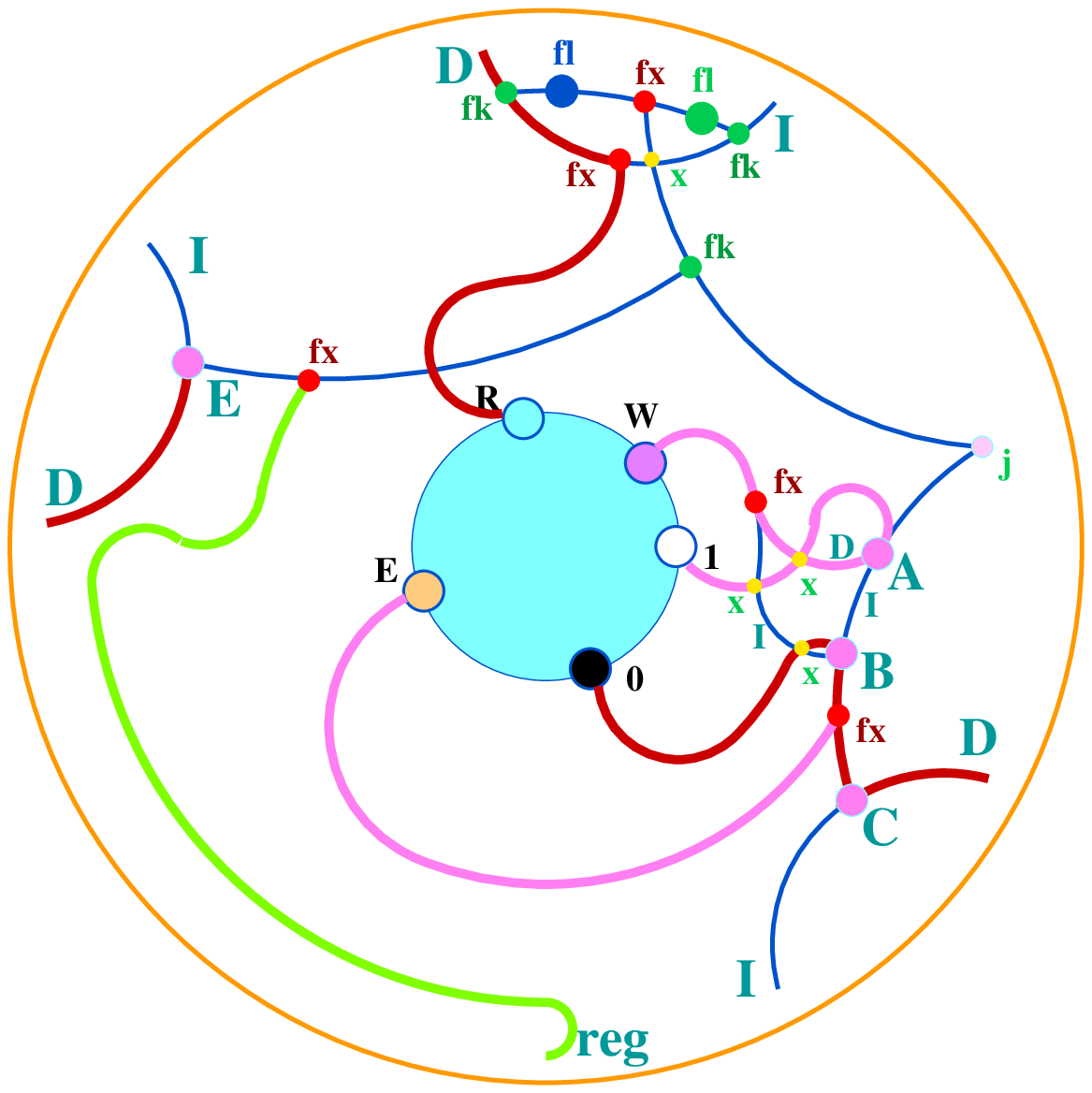}
\hfill}
\vspace{-25pt}
\begin{fig}\label{f_uDDTb}
\leurre
Inside \DDT, to left, to right, the tracks followed by an incrementing, decrementing respectively 
locomotive.
\end{fig}
}

Figure~\ref{f_uDDTb} illustrates the tracks followed by a locomotive depending on whether it has
to increment of to decrement the register. On both pictures, the red tracks are followed when
the bit contained in the memory is that attached to the type of instruction. Otherwise, the tracks
are in light mauve. In both cases, the mauve tracks lead to the \WW-gate so that the bit~$b$ is 
changed into \hbox{$1-b$}. Accordingly, in that case the locomotive exit the memory through 
its \EE-gate and goes to $C$ via a passive fixed switch and from~$C$ it goes to the register in
order to perform the appropriate operation.

\label{reg_op}
{\bf \ref{reg_op}.d\ Decrementing and incrementing the register}

The operation performed by a locomotive inside the register is illustrated by Figure~\ref{f_reg_op}.
The figure illustrates a standard situation for a non trivial value of the content: the \MM-tile is
at some distance from the beginning of the structure, depending on the initial content of the 
register which may be empty. On the leftmost picture of the figure, we can see the
\YY-cells 1(6), 1(7) and 1(1) around the \MM-cell at~0 which we call the {\bf hat} of the register.
In the rightmost picture of the figure we also see the hat around the \MM-cell at 1(3), it consists 
of the \YY-cells 1(4), 0 and 1(2). In the third picture, illustrating the incrementing operation,
the hat consists of four \YY-cells around the \MM-cell at 1(7): 2(7), 3(7), 4(7) and 2(1). 

Figure~\ref{f_reg_op} illustrates how the operation is performed. The figure is split into two
parts. To left, the register before the operation, to right, the register after it was performed.
In that right-hand side part, two possibilities, to left: incrementing; to right, decrementing
as mentioned.

\vtop{
\ligne{\hfill
\includegraphics[scale=1]{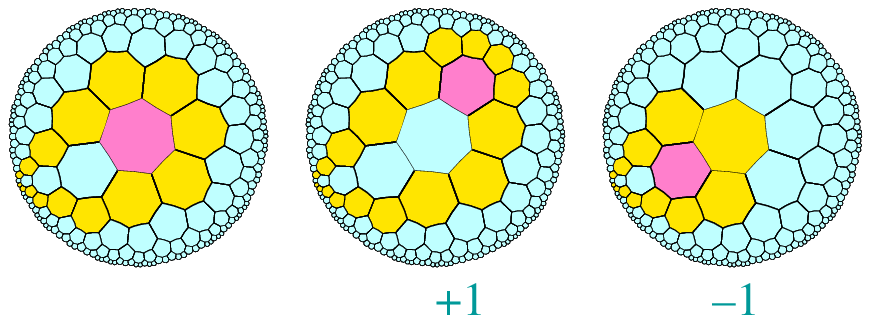}
\hfill}
\vspace{-10pt}
\begin{fig}\label{f_reg_op}
\leurre
To left, the register before the operation. To right, the configuration after the operation. 
\end{fig}
}

The action illustrated by Figure~\ref{f_reg_op} has been tested by a computer program.

\section{Rules}\label{srules}

   The definition of the rules will give us the opportunity to more explicitly study the behaviour
of the automaton.

   We start the section by introducing a formalism for writing schemes of rules which will
allow us to reduce the number of rules. Then, in each further sub-sections,
we successively examine the cases studied in the sub-subsections of 
Subsection~\ref{newrailway}. 

\subsection{Fixing notations}\label{sbnot}

As far as we assume the rules to be rotation invariant, we gather the rules having the same
neighbourhood up to rotation by taking the least one according to the lexicographic
order we set on the states which is the following: \hbox{\WW, \YY, \BB, \GG, \RR, \MM}. 
From now on, as far as we speak about the cellular automaton, we shall say cell in
place of tiles. As an example, the central tile of a window will become the central
cell of the window. The reason is that a cell is supported by a tile but it also contains
the finite automaton, the same for all cells, which rules the change of cell at each tip
of the discrete clock. All those conditions are just an application of the definition of
a cellular automaton. 
The first letter is the current state of the cell and the
last one is the new cell obtained from the current state of the cell and the current 
states of its neighbours: if we order the neighbours according to the letters from~1 up
to~7 from left to right in the word delimited between two commas, the $i^{\rm th}$
letter gives the state of the corresponding $i^{\rm th}$ neighbour of the cell. The first
neighbour is, in that case, depending on the neighbourhood. Now, from rotation invariance,
we consider the smallest word obtained from the states of the neighbours, so that
if at least two states are different, the first neighbour is uniquely defined in that
way.

\def\schrule #1 #2 #3 {\hbox{\hbox{\ftt{#1}-\hbox{\ftt#2}:\hbox{\ftt#3}\hfill}}
}

Take, as an example the motion rule on a segment of line for a blue locomotive, namely 
\schrule {Y} {WWWBWWY} {B}. It is the minimal form of rule~150{} in Table~\ref{t_rules}. In the
table, rule~150 is displayed as \schrule {Y} {BWWYWWW} {B}. The rule replaces the following ones:
\vskip 5pt
\vtop{
\ligne{\hfill
\vtop{\leftskip 0pt\parindent 0pt\hsize=61pt
\ligne{{$a$} \schrule {Y} {WWWYWWY} {Y} \hfill}
\ligne{{$b$} \schrule {Y} {WWWBWWY} {B} \hfill}
\ligne{{$c$} \schrule {Y} {WWBWWYW} {B} \hfill}
\ligne{{$d$} \schrule {Y} {WBWWYWW} {B} \hfill}
}
\hfill
\vtop{\leftskip 0pt\parindent 0pt\hsize=61pt
\ligne{{$e$} \schrule {Y} {BWWYWWW} {B} \hfill}
\ligne{{$f$} \schrule {Y} {WWYWWWB} {B} \hfill}
\ligne{{$g$} \schrule {Y} {WYWWWBW} {B} \hfill}
\ligne{{$h$} \schrule {Y} {YWWWBWW} {B} \hfill}
}
\hfill}
}
\vskip 5pt
\vskip 5pt
Say that a rule is {\bf conservative} if and only if the new state is the same as the current one
and also if the states of the neighbours are also unchanged when the rules is applied. If
the state of the cell only is unchanged, we speak of a {\bf witness rule}. As an example,
rule~7, namely \schrule {W} {WWWWWBR} {W} is a witness rule as far as the \BB- and \RR-cells in
the neighbourhood of the current cell to which the rule applies belong to a blue locomotive.

\subsection{Rules for the tracks}\label{sbtracks}
The rules for the motion of a locomotive applying to the cells of a track are given in 
Table~\ref{t_move}.

\vspace{-10pt}
\vtop{\leftskip 0pt\parindent 0pt
\begin{tab}\label{t_move}
\leurre
Motion rules of a locomotive on the tracks.
\end{tab}
\vspace{-5pt}
\ligne{\hfill
\vtop{\leftskip 0pt\parindent 0pt\hsize=61pt
\ligne{\schrule {Y} {WWWBWWY} {B} {\ftt{152}}\hfill}
\ligne{\schrule {B} {WWWRWWY} {R} {\ftt{17}}\hfill}
\ligne{\schrule {R} {WWWYWWB} {Y} {\ftt{18}}\hfill}
\ligne{\schrule {Y} {WWWYWWR} {Y} {\ftt{21}}\hfill}
}
\hfill
\vtop{\leftskip 0pt\parindent 0pt\hsize=61pt
\ligne{\schrule {Y} {WWWGWWY} {G} {\ftt{161}}\hfill}
\ligne{\schrule {G} {WWWRWWY} {R} {\ftt{24}}\hfill}
\ligne{\schrule {R} {WWWYWWG} {Y} {\ftt{25}}\hfill}
\ligne{\schrule {Y} {WWWYWWR} {Y} {\ftt{21}}\hfill}
}
\hfill
\vtop{\leftskip 0pt\parindent 0pt\hsize=61pt
\ligne{\schrule {Y} {WWWYWWB} {B} {\ftt{64}}\hfill}
\ligne{\schrule {B} {WWWYWWR} {R} {\ftt{53}}\hfill}
\ligne{\schrule {R} {WWWBWWY} {Y} {\ftt{60}}\hfill}
\ligne{\schrule {Y} {WWWRWWY} {Y} {\ftt{62}}\hfill}
}
\hfill
\vtop{\leftskip 0pt\parindent 0pt\hsize=61pt
\ligne{\schrule {Y} {WWWYWWG} {G} {\ftt{94}}\hfill}
\ligne{\schrule {G} {WWWYWWR} {R} {\ftt{88}}\hfill}
\ligne{\schrule {R} {WWWGWWY} {Y} {\ftt{91}}\hfill}
\ligne{\schrule {Y} {WWWRWWY} {Y} {\ftt{62}}\hfill}
}
\hfill}
}
\vskip 5pt

In the above rules, the two left-hand side columns apply to a locomotive on a segment of line
or on a circular arc under a counter-clockwise motion. The two-right-hand side columns apply to 
a locomotive on a circular arc under a clockwise motion. Note that rules~21 and~62 occur twice
as far as they concern a cell of the track witnessing the rear of a leaving locomotive so that such
a rule apply to any locomotive, whatever its colour is. Also, rules~152 and~161 have the same 
neighbourhood but different current states; a similar remark holds for rules~18 and~64, also for
rules~161 and~91 and also for rules~25 and~94. The rules for a \GG-locomotive are obtained from 
those for a \BB-one by replacing the state \BB- by \GG- in the rules for a \BB-locomotive.

\def\FF{{\tt F}}
A track along an arc of a circle requires additional rules. They have the form
\schrule {Y} {WWWWFWY} {F} , where \FF{} stands for \BB{} or \GG, the front of a locomotive.
We have the rules \schrule {Y} {WWWWFWY} {F}, the rule \schrule {F} {WWWWRWY} {R}, the rule 
\schrule {R} {WWWWYWF} {Y} and the rule \schrule {Y} {WWWWYWR} {Y}. They are rules 36, 38, 54
and 42 respectively for a \BB-locomotive, rules 75, 77, 79 and 42 again respectively for a \GG-%
locomotive. All those rules are used for a clockwise motion. The rules for a counter-clockwise 
motion are obtained from the clockwise rules by exchanging the states between the fifth and the 
seventh neighbour. As an example we get the rules \schrule {Y} {WWWWYWF} {F} from the rules
\schrule {Y} {WWWWFWY} {F}, rules 54, 89 for a \BB-,\GG-locomotive respectively.

Figure~\ref{f_voies} shows us the idle configurations of cells of the tracks where along which a
clockwise motion is reversed  into a counter-clockwise one and conversely. Figure~\ref{f_m_voies}
illustrates the motion of a \BB-locomotive along the tracks illustrated by Figure~\ref{f_voies}
in both direction. The reader may check the rules of Table~\ref{t_move} on Figure~\ref{f_m_voies}
which works also for a \GG-locomotive, replacing the \BB-tiles by \GG-ones. 

\vskip 5pt
\vtop{
\ligne{\hfill
\includegraphics[scale=0.5]{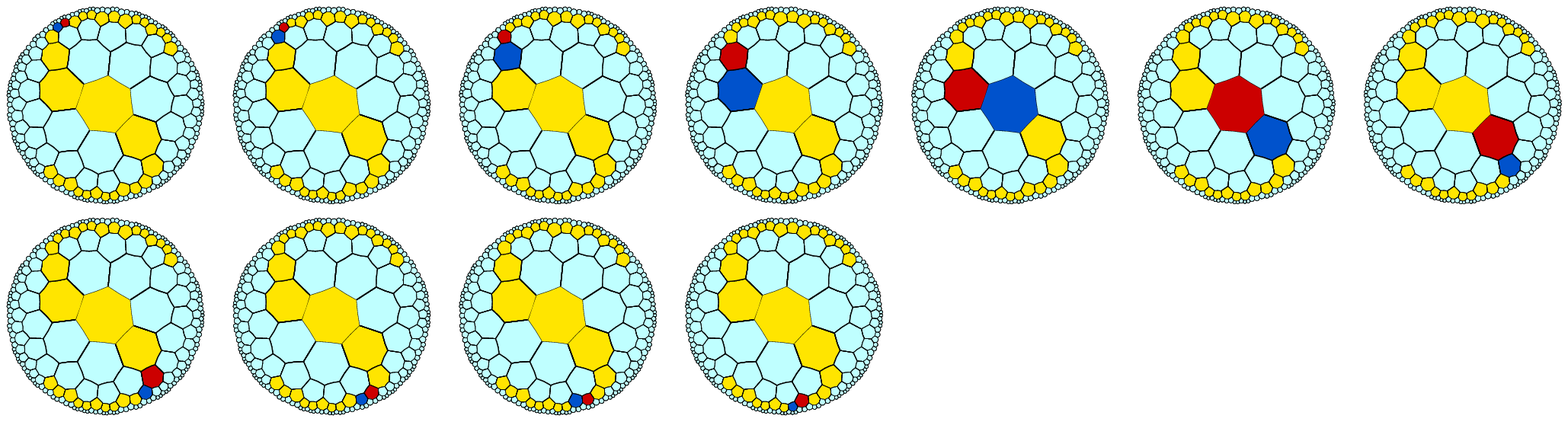}
\hfill}
\vspace{-10pt}
\ligne{\hfill
\includegraphics[scale=0.5]{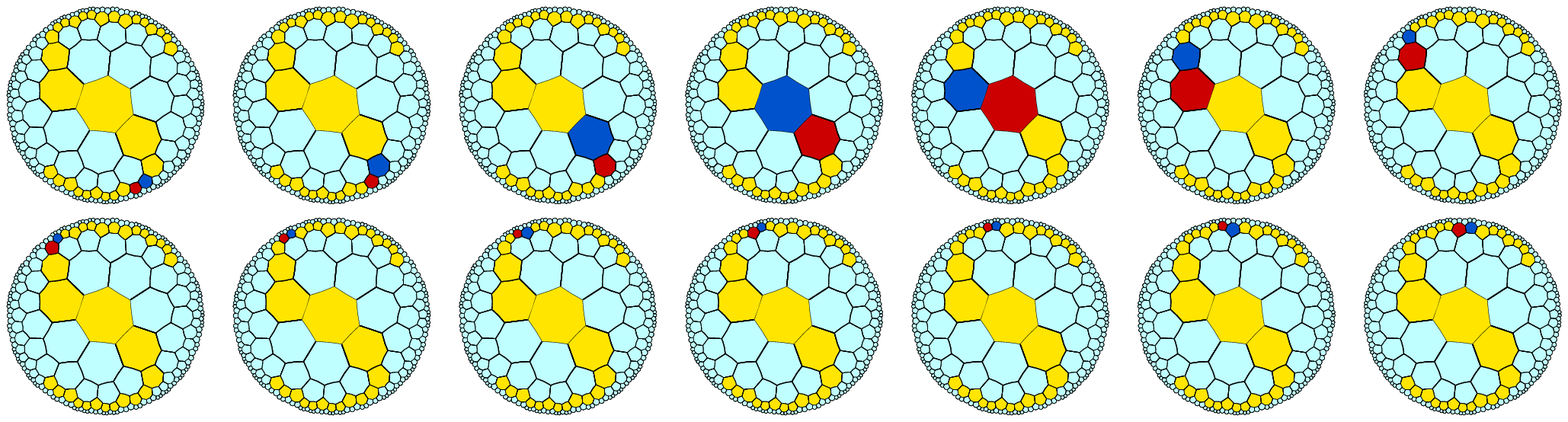}
\hfill}
\begin{fig}\label{f_m_voies}
\leurre
Top rows: motion of a blue locomotive from top to bottom. Bottom rows: motion of a blue locomotive
from bottom to top. Note that on the arcs of a circle we have both a clockwise and a 
counter-clockwise motion.
\end{fig}
}
\vskip 5pt

\subsection{Fixed switch and fork}\label{sbfxfk}

We start with the fixed switch, considering its passive version only.

We have four cases: a blue locomotive coming from the left-hand, right-hand side branch and 
the similar two sub-cases with a green locomotive.
We refer the reader to Figures~\ref{f_i_fix} and  \ref{f_auxil} to follow the application of the 
rules given by Table~\ref{tfx} where the notations are those introduced in 
Subsection~\ref{sbnot}. We also use the same meta-symbol \FF{} to replace \BB{} and \GG.
Figure~\ref{f_m_fix} illustrates the motion of a locomotive passively arriving to the switch.
\vskip 5pt
\vtop{
\ligne{\hfill
\includegraphics[scale=0.5]{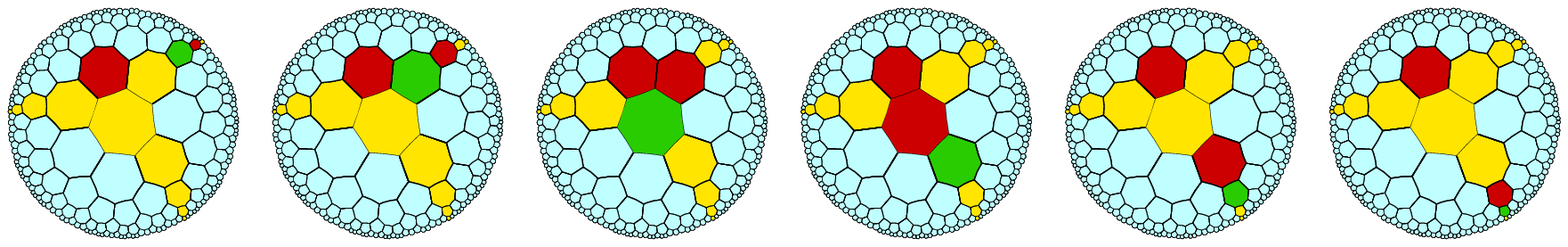}
\hfill}
\ligne{\hfill
\includegraphics[scale=0.5]{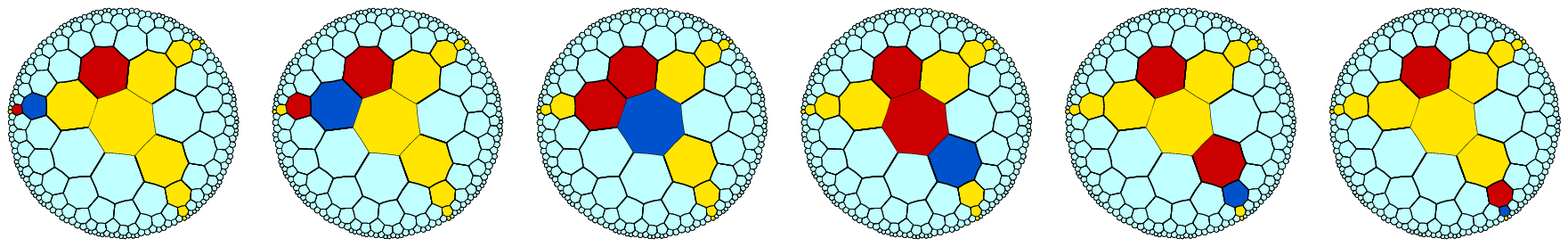}
\hfill}
\begin{fig}\label{f_m_fix}
\leurre
The motion of a locomotive passively crossing a fixed switch.
Top, a \GG-locomotive arriving from the right. Bottom, a \BB-locomotive arriving from the left.
The opposite motions for a \BB- or a \GG-locomotive are easily derived from the pictures.
\end{fig}
}
\vskip 5pt
\def\fts#1{{\small #1}}
\vtop{
\begin{tab}\label{tfx}
\leurre
Rules for the fixed switch.
\end{tab}
\vskip-7pt
\ligne{\hfill conservative rules for the fixed switch\hfill}
\vskip-2pt
\ligne{\hfill
\vtop{\leftskip 0pt\parindent 0pt\hsize=91pt
\vskip-2pt
\ligne{\hfill at 0, then 1(2) and then 1(7)\hfill}
\ligne{\hfill \schrule {Y} {WWYWYRY} {Y} \hfill {\ftt {119}}\hfill}
\ligne{\hfill \schrule {Y} {WWYWWYR} {Y} \hfill {\ftt {135}}\hfill}
\ligne{\hfill \schrule {Y} {WWYWWRY} {Y} \hfill {\ftt {122}}\hfill}
}
\hfill}
\ligne{\hfill motion rules for the fixed switch\hfill}
\ligne{\hfill\footnotesize 1$^{st}$, 2$^{nd}$ number: \FF = \BB, \GG{} resp.\hfill}
\ligne{\hfill
\vtop{\leftskip 0pt\parindent 0pt\hsize=100pt
\ligne{\hfill at cell 0 from left\hfill}
\ligne{\schrule {Y} {WWYWYRY} {Y} \hfill {\ftt {119}}\hfill}
\ligne{\schrule {Y} {WWYWYRF} {F} \hfill {\ftt {123,137}}\hfill}
\ligne{\schrule {F} {WWYWYRR} {R} \hfill {\ftt {126,140}}\hfill}
\ligne{\schrule {R} {WWFWYRY} {Y} \hfill {\ftt {131,144}}\hfill}
\ligne{\schrule {Y} {WWRWYRY} {Y} \hfill {\ftt {134}}\hfill}
}
\hfill
\vtop{\leftskip 0pt\parindent 0pt\hsize=100pt
\ligne{\hfill at cell 0 from right\hfill}
\ligne{\schrule {Y} {WWYWYRY} {Y} \hfill {\ftt {117}}\hfill}
\ligne{\schrule {Y} {WWYWFRY} {F} \hfill {\ftt {146,155}}\hfill}
\ligne{\schrule {F} {WWYWRRY} {R} \hfill {\ftt {149,158}}\hfill}
\ligne{\schrule {R} {WWFWYRY} {Y} \hfill {\ftt {129,144}}\hfill}
\ligne{\schrule {Y} {WWRWYRY} {Y} \hfill {\ftt {134}}\hfill}
}
\hfill}
\vskip 2pt
\ligne{\hfill
\vtop{\leftskip 0pt\parindent 0pt\hsize=100pt
\ligne{\hfill at cell 1(2)\hfill}
\ligne{\hfill\schrule {Y} {WWYWWYR} {Y} \hfill {\ftt {135}}\hfill} 
\ligne{\hfill\schrule {Y} {WWFWWYR} {F} \hfill {\ftt {121,136}}\hfill} 
\ligne{\hfill\schrule {F} {WWRWWYR} {R} \hfill {\ftt {125,139}}\hfill} 
\ligne{\hfill\schrule {R} {WWYWWFR} {Y} \hfill {\ftt {128,142}}\hfill} 
\ligne{\hfill\schrule {Y} {WWYWWRR} {Y} \hfill {\ftt {133}}\hfill} 
}
\hfill
\vtop{\leftskip 0pt\parindent 0pt\hsize=100pt
\ligne{\hfill at cell 1(7)\hfill}
\ligne{\hfill\schrule {Y} {WWYWWRY} {Y} \hfill {\ftt {122}}\hfill} 
\ligne{\hfill\schrule {Y} {WWFWWRY} {F} \hfill {\ftt {145,154}}\hfill} 
\ligne{\hfill\schrule {F} {WWRWWRY} {R} \hfill {\ftt {148,157}}\hfill} 
\ligne{\hfill\schrule {R} {WWYWWRF} {Y} \hfill {\ftt {153,162}}\hfill} 
\ligne{\hfill\schrule {Y} {WWYWWRR} {Y} \hfill {\ftt {133}}\hfill} 
}
\hfill}
}
\vskip 10pt
Table~\ref{tfx} gives the rules used for the cells~0, 1(1) and 1(7). The rules for cell~0 are
witness rules rule~119 being excepted as far as it is conservative. For the cells~1(2) and 1(7) 
rules 135 and 122 are conservative, the other are motion rules. Note that the same rule~133 is
used for both branches when the cell witnesses a leaving rear of the locomotive. Also note that
the rules when the front of the locomotive is in 1(5) are different for 1(2) and for 1(7): 
128,142 on the left-hand side branch, 153,162 on the right-hand side one. In the previous couples
of numbers of rules, the first, second number deals with \BB-, \GG-locomotives respectively.
The application of those rules can be checked on Figure~\ref{f_m_fix}.

\vtop{
\begin{tab}\label{tfk}
\leurre
Rules for the fork.
\end{tab}
\vskip-7pt
\ligne{\hfill conservative rules for the fork\hfill}
\vskip-2pt
\ligne{\hfill
\vtop{\leftskip 0pt\parindent 0pt\hsize=91pt
\vskip-2pt
\ligne{\hfill at 0 and then 1(1)\hfill}
\ligne{\schrule {Y} {WWYWYWY} {Y} \hfill {\ftt {98}}\hfill}
\ligne{\schrule {W} {WWWWYYY} {W} \hfill {\ftt {56}}\hfill}
}
\hfill}
\vskip 2pt
\ligne{\hfill motion rules and witness rules for the fork\hfill}
\ligne{\hfill
\vtop{\leftskip 0pt\parindent 0pt\hsize=100pt
\ligne{\hfill at cell 0\hfill}
\ligne{\schrule {Y} {WWYWYWY} {Y} \hfill {\ftt {98}}\hfill}
\ligne{\schrule {Y} {WWFWYWY} {F} \hfill {\ftt {99,111}}\hfill}
\ligne{\schrule {F} {WWRWYWY} {R} \hfill {\ftt {101,113}}\hfill}
\ligne{\schrule {R} {WWYWFWF} {Y} \hfill {\ftt {104,116}}\hfill}
\ligne{\schrule {Y} {WWYWRWR} {Y} \hfill {\ftt {107}}\hfill}
}
\hfill
\vtop{\leftskip 0pt\parindent 0pt\hsize=100pt
\ligne{\hfill at cell 1(1)\hfill}
\ligne{\schrule {W} {WWWWYYY} {W} \hfill {\ftt {56}}\hfill}
\ligne{\schrule {W} {WWWWYFY} {W} \hfill {\ftt {102,114}}\hfill}
\ligne{\schrule {W} {WWWWFRF} {W} \hfill {\ftt {105,117}}\hfill}
\ligne{\schrule {W} {WWWWRYR} {W} \hfill {\ftt {108}}\hfill}
}
\hfill}
}
\vskip 10pt
Note rules~104,116 and 107 and also 105,117{} in which we can see the occurrence of two fronts
of the duplicated locomotives. Also note that the rule where the cell witnesses a leaving 
locomotive is the same for both types of locomotive. The rules of the table can be checked on
Figure~\ref{f_m_fk} which illustrates the motion of a locomotive through a fork.

\vskip 5pt
\vtop{
\ligne{\hfill
\includegraphics[scale=0.5]{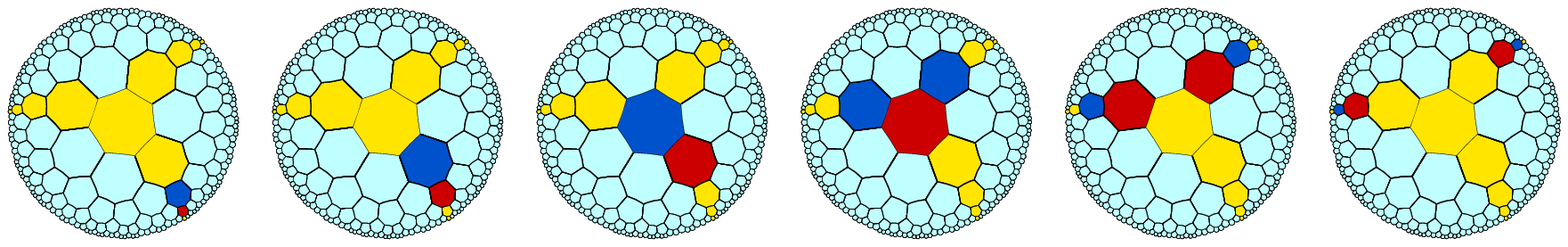}
\hfill}
\ligne{\hfill
\includegraphics[scale=0.5]{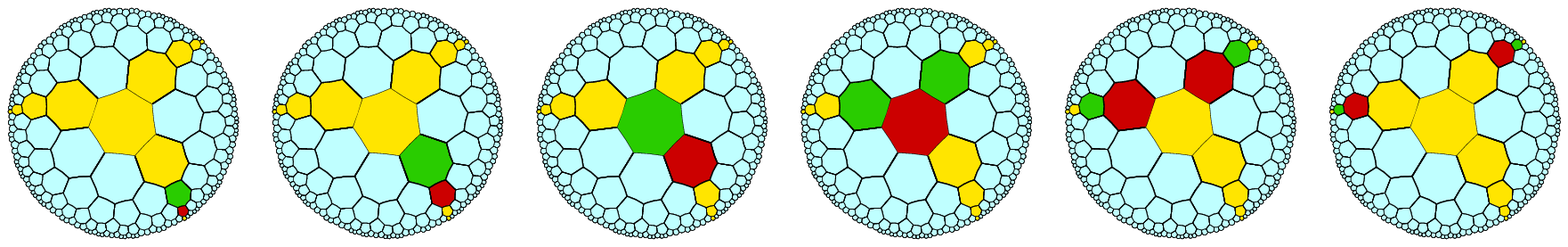}
\hfill}
\begin{fig}\label{f_m_fk}
\leurre
Top: duplication of a \BB-locomotive. Bottom: duplication of a \GG-one.
\end{fig}
}

\subsection{Changers and filters}\label{schflts}

   Below, two tables give the rules managing the changers and the filters and for that latter
structure we also indicate the rules for changing the colour of the filter. Table~\ref{tchs}
gives the rules concerning the changers and Table~\ref{tflts} gives those ruling the filters.

\def\HH{{\tt H}}
In the following tables, the occurrences of \FF{} always stands for \BB{} or always stands 
for \GG, while \HH{} always stands for \GG{} or for \BB{} respectively. The same convention as in 
Tables~\ref{tfx} and~\ref{tfk} holds for the couples of numbers for the rules.

\vtop{
\begin{tab}\label{tchs}
\leurre
Rules for the changers. First, the blue one and then the green one. In both of them cell~0, and the
cells~1(1) and 1(7) are considered. 
\end{tab}
\vskip-7pt
\ligne{\hskip 10pt the blue changer\hfill\hfill}
\vskip -2Pt
\ligne{\hfill
\vtop{\leftskip 0pt\parindent 0pt\hsize=85pt
\vskip 2pt
\ligne{\hfill\fts{cell 0}\hfill}
\vskip-2pt
\ligne{\schrule {Y} {WWWYFWY} {Y} {\ftt{180,163}}\hfill}
\ligne{\schrule {Y} {WWWFFWY} {F} {\ftt{185,168}}\hfill}
\ligne{\schrule {F} {WWWRFWY} {R} {\ftt{189,172}}\hfill}
\ligne{\schrule {R} {WWWYFWF} {Y} {\ftt{192,175}}\hfill}
\ligne{\schrule {Y} {WWWYFWR} {Y} {\ftt{194,177}}\hfill}
}
\vtop{\leftskip 0pt\parindent 0pt\hsize=85pt
\vskip 2pt
\ligne{\hfill\fts{cell 1(1)}\hfill}
\ligne{\schrule {F} {WWYYWYY} {F} {\ftt{181,164}}\hfill}
\ligne{\schrule {F} {WWYFWYY} {F} {\ftt{186,169}}\hfill}
\ligne{\schrule {F} {WWFRWYY} {F} {\ftt{190,173}}\hfill}
\ligne{\schrule {F} {WWRYWYY} {F} {\ftt{193,176}}\hfill}
}
\vtop{\leftskip 0pt\parindent 0pt\hsize=85pt
\vskip 2pt
\ligne{\hfill\fts{cell 1(7)}\hfill}
\ligne{\schrule {Y} {WWYWWFY} {Y} {\ftt{196,179}}\hfill}
\ligne{\schrule {Y} {WWBYWWG} {B} {\ftt{184}}\hfill}
\ligne{\schrule {Y} {WWBWWGY} {G} {\ftt{167}}\hfill}
\ligne{\schrule {F} {WWFYWWR} {R} {\ftt{188,171}}\hfill}
\ligne{\schrule {R} {WWYWWFF} {Y} {\ftt{191,174}}\hfill}
\ligne{\schrule {Y} {WWYWWFR} {Y} {\ftt{151,160}}\hfill}
}
\hfill}
}
\vskip 10pt

Note that rule~167 is obtained from rule~184 by exchanging the states \BB- and \GG-, but
that change modifies the alphabetic order of the corresponding minimal rule. The rules can be
checked on Figure~\ref{f_m_ch}.

\vtop{
\begin{tab}\label{tflts}
\leurre
Rules for the filters. Here too, we use the formalism allowing us to present both filters with 
the same meta-rules: \FF{} and \HH{} stand both for \BB- or \GG- with the convention that \FF{} and
\HH{} are always different when they occur in the same rule and in the same group of rules.
The same convention as in Tables~\ref{tfx} and~\ref{tfk} holds for the couples of numbers for 
the rules.
\end{tab}
\vskip-7pt
\ligne{\hfill the working of the filter\hfill\hfill}
\vskip -5pt
\ligne{\hfill
\vtop{\leftskip 0pt\parindent 0pt\hsize=85pt
\vskip 2pt
\ligne{\hfill\fts{cell 0}\hfill}
\ligne{\schrule {F} {WYYWRYR} {F} {\ftt{197,220}}\hfill}
\ligne{\schrule {F} {WYFWRYR} {F} {\ftt{204,227}}\hfill}
\ligne{\schrule {F} {WFRWRYR} {F} {\ftt{207,230}}\hfill}
\ligne{\schrule {F} {WRYWRYR} {F} {\ftt{211,234}}\hfill}
}
\hfill
\vtop{\leftskip 0pt\parindent 0pt\hsize=85pt
\vskip 2pt
\ligne{\hfill\fts{cell 1(5)}\hfill}
\ligne{\schrule {Y} {WWWYWFY} {Y} {\ftt{215,238}}\hfill}
\ligne{\schrule {Y} {WWWFWFY} {F} {\ftt{201,224}}\hfill}
\ligne{\schrule {Y} {WWWHWFY} {Y} {\ftt{216,237}}\hfill}
\ligne{\schrule {F} {WWWRWFY} {R} {\ftt{205,228}}\hfill}
\ligne{\schrule {R} {WWWYWFF} {Y} {\ftt{209,230}}\hfill}
\ligne{\schrule {Y} {WWWYWFR} {Y} {\ftt{213,234}}\hfill}
}
\hfill}
\vskip 2pt
\ligne{\hfill changing the filter\hfill} 
\vskip -5pt
\ligne{\hfill
\vtop{\leftskip 0pt\parindent 0pt\hsize=85pt
\vskip 2pt
\ligne{\hfill\fts{cell 1(1)}\hfill}
\ligne{\schrule {Y} {WWRFRWY} {Y} {\ftt{195,218}}\hfill}
\ligne{\schrule {Y} {WWWFWFY} {F} {\ftt{199,222}}\hfill}
\ligne{\schrule {Y} {WWWHWFY} {Y} {\ftt{214,239}}\hfill}
\ligne{\schrule {F} {WWWRWFY} {R} {\ftt{203,228}}\hfill}
\ligne{\schrule {R} {WWWYWFF} {Y} {\ftt{207,232}}\hfill}
\ligne{\schrule {Y} {WWWYWFR} {Y} {\ftt{211,236}}\hfill}
}
\hfill
\vtop{\leftskip 0pt\parindent 0pt\hsize=85pt
\vskip 2pt
\ligne{\hfill\fts{cell 0}\hfill}
\ligne{\schrule {F} {WYYWRYR} {F} {\ftt{197,220}}\hfill}
\ligne{\schrule {F} {WYYWRHR} {H} {\ftt{245,251}}\hfill}
\ligne{\schrule {H} {WYYWRYR} {H} {\ftt{220,197}}\hfill}
}
\hfill}
}
\vskip 10pt
\vtop{
\ligne{\hfill
\includegraphics[scale=0.5]{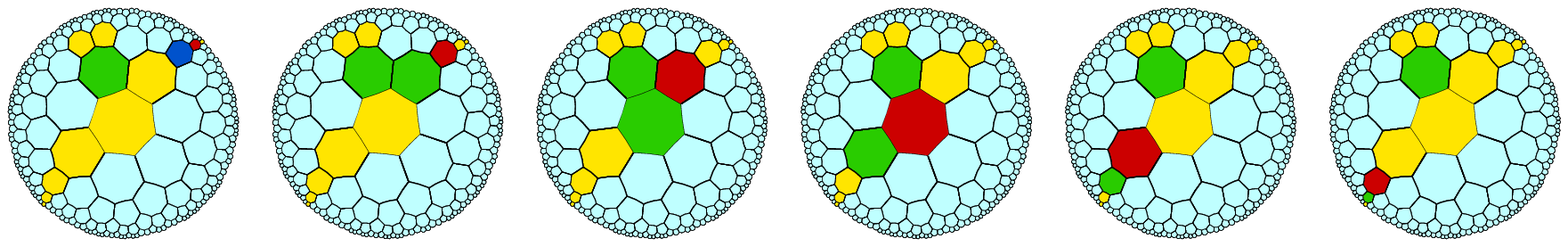}
\hfill}
\vspace{-10pt}
\ligne{\hfill
\includegraphics[scale=0.5]{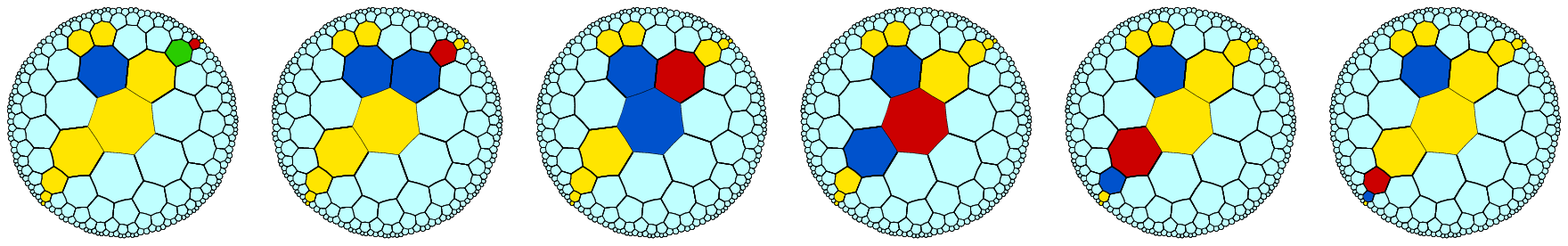}
\hfill}
\vspace{-30pt}
\begin{fig}\label{f_m_ch}
\leurre
Top: changing a \BB-locomotive to a \GG-one. Bottom: changing a \GG-locomotive to a \BB-one.
\end{fig}
}

Figure~\ref{f_m_ch} shows us the action of the changer on a locomotive of the opposite colour: that
latter one takes the colour of the changer.

Figure~\ref{f_m_flt} similarly shows us the action of the filter. We can see that a filter let a
locomotive of the same colour go while it destroys a locomotive of the opposite colour.

As far as we have often used filters in the implementations of the switches, we know that it was
needed to program a filter by giving it the possibility to change its colour. The appropriate rules
of Table~\ref{tflts} allow a suitable locomotive to do that. Figure~\ref{f_chflt} allows the reader
to check those rules. That latter figure also explains us why Figure~\ref{f_auxil} has two \YY-cells
close to the cell at~0 whose colour defines that of the filter and who stand at the opposite of 
the \YY-cells of the track on which the locomotive to be controlled passes. Those \YY-cells are the
end of a track arriving to the filter in order to change its colour as dictated by rules~245 
and~251, namely \schrule {B} {WYYWRGR} {G} {} and \schrule {G} {WYYWRBR} {B} {} respectively.

\vtop{
\ligne{\hfill
\includegraphics[scale=0.5]{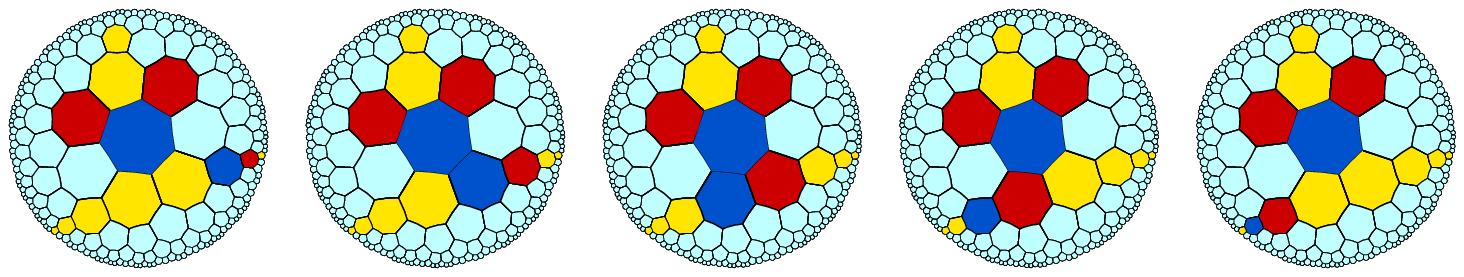}
\hfill}
\ligne{\hfill
\includegraphics[scale=0.5]{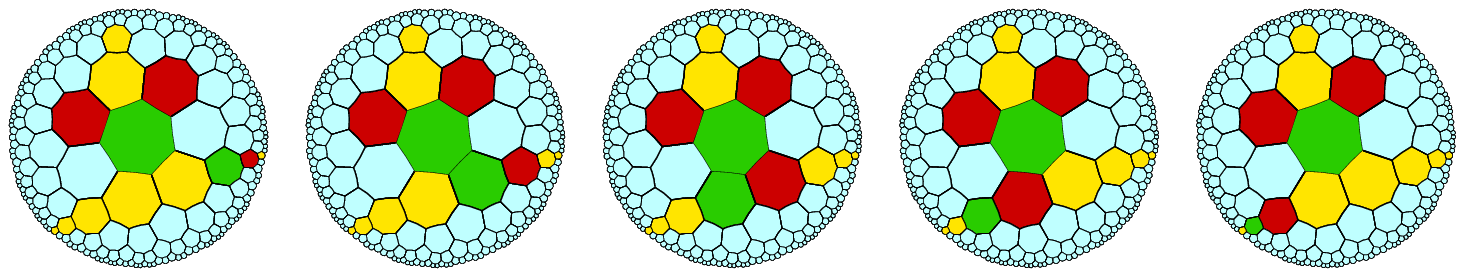}
\hfill}
\ligne{\hfill
\includegraphics[scale=0.5]{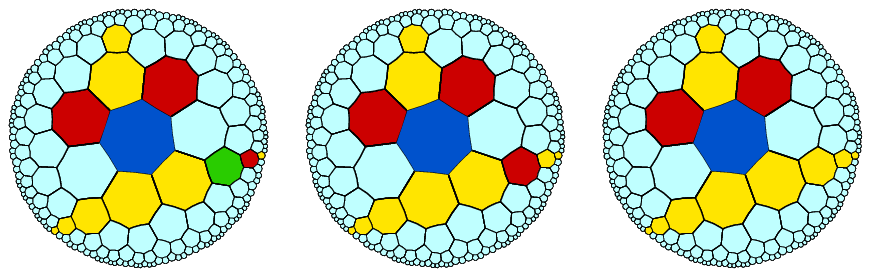}
\includegraphics[scale=0.5]{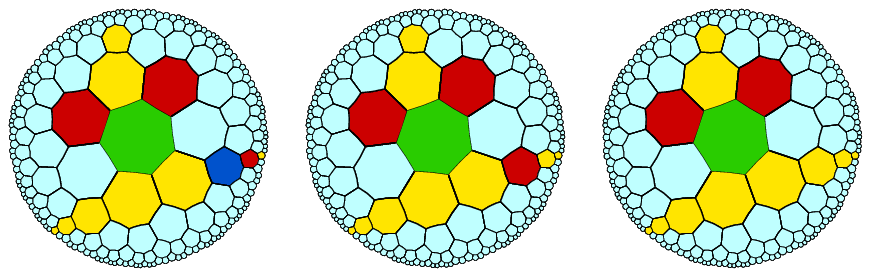}
\hfill}
\begin{fig}\label{f_m_flt}
\leurre
Top: when the filter let the locomotive of its colour go. Bottom, the filter stops the locomotive
of the opposite colour.
\end{fig}
}

\vtop{
\ligne{\hfill
\includegraphics[scale=0.5]{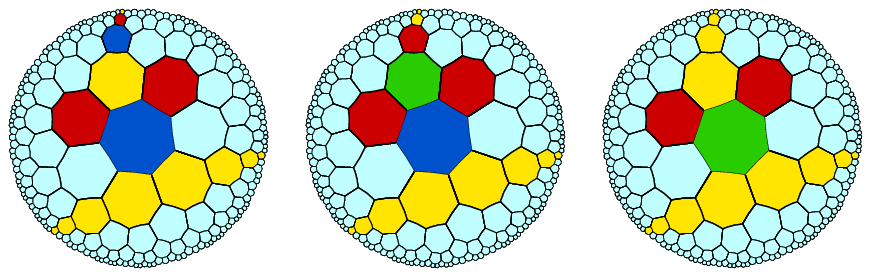}
\includegraphics[scale=0.5]{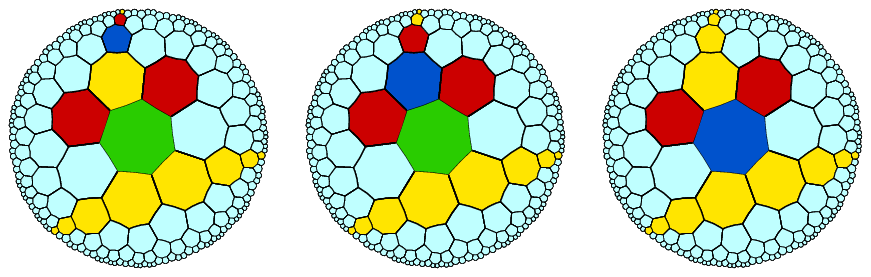}
\hfill}
\begin{fig}\label{f_chflt}
\leurre
Changing the colour of the filter.
\end{fig}
}

\def\reg #1 {\hbox{$\mathcal R$(#1)}}
\subsection{Registers}~\label{sbregs}

   Let $\mathcal R$ denote a register. The cells belonging to $\mathcal R$ are denoted by 
\reg{n}{} with $n\in\mathbb N$. The register is empty when $n=0$. If $n$ is the value of the
content of the register, the cell \reg{n} {} is \MM. Accordingly, when the register is empty,
\reg{0} {} is \MM. As can be seen from Figure~\ref{f_reg}, the transformation from \reg{n} {} to
\reg{n+1} {} or \reg{n-1} {} entails much work. About that transformation, we distinguish the case
when $n=0$ from that when $n>0$ and, in the latter case for a decrementing instruction we 
distinguish the case $n=1$ from the case $n>1$.

   The case when $n=0$ requires a special configuration for \reg{0} {} in order the decrementing
operation should not be processed. Instead the \GG-locomotive should go along a specific track
as mentioned in Subsection~\ref{sbbregdisc}. Now, when a locomotive go to \reg{n}, when $n>1$ in 
order to perform an operation, it must cross \reg{0}. Accordingly, there are two possible tracks
followed by a locomotive leaving \reg{0}, depending on whether it could not perform the operation
or it could do it as far as it has to increment $\mathcal R$ or it has to operate on \reg{n} for
$n>0$. Figure\ref{f_debreg} illustrates the situation around \reg{0}. There are two tracks for a 
locomotive leaving the neighbourhood of \reg{0}: one follows the cells 1(4), 1(5), 1(6), 2(7),
7(7), 20(7) and further towards \reg{n}; the other track follows the cells 1(4), 1(5), 2(6),
7(6), 20(6) and further towards the \DDT{} structure attached to $\mathcal R$.

\vtop{
\ligne{\hfill
\includegraphics[scale=2]{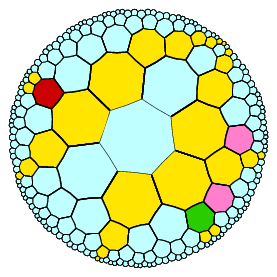}
\hfill}
\begin{fig}\label{f_debreg}
\leurre
The idle configuration of \reg{0}. Note that cell~0 is \MM. Note the other \MM-cells.
\end{fig}
}

   The bifurcation occurs at 1(5): it concerns two other cells, 2(6) and 1(6). It is the reason of
two \MM-cells, one at 4(5), which is seen from both 1(5) and 2(6) and the other sits at 3(6) which
is seen from both 1(6) and 2(6). The control is easy for a \BB-locomotive which has to increment
the register. The situation is less simple for a \GG-locomotive: if incrementing fails, the 
locomotive has to go to 2(6), otherwise it has to go to 1(6). The distinction is also visible at
cell~0: it is \MM{} if and only if and only if the content of the register is 0. The difference
is shown by Figures~\ref{f_m_debreg} and by \ref{f_reg_z}. The concerned rules are given by 
Table~\ref{t_debreg} only deals with the case when the locomotive crosses \reg{0} {} and it is
assumed that $n\geq1$. The case $n=0$ is dealt with separately. It is illustrated by 
Figure~\ref{f_reg_z} and some rules are given by Table~\ref{t_z_op}. The
case $n=1$ is dealt with Figure~\ref{f_decun} and rules are given by Table\ref{t_u_op}. 

\vtop{
\begin{tab}\label{t_debreg}
\leurre
The rules managing the crossing of \reg{0}, general case.
\end{tab}
\ligne{\hfill incrementing, $n>0$\hfill}
\ligne{\hfill
\vtop{\leftskip 0pt\parindent 0pt\hsize=85pt
\ligne{\hfill cell 0\hfill}
\ligne{\schrule {W} {WYYWYYY} {W} {\ftt{274}} \hfill}
\ligne{\schrule {W} {WYYWBYY} {W} {\ftt{372}} \hfill}
\ligne{\schrule {W} {WYYWRBY} {W} {\ftt{371}} \hfill}
\ligne{\schrule {W} {WYYWYRB} {W} {\ftt{373}} \hfill}
\ligne{\schrule {W} {WYYWYYR} {W} {\ftt{273}} \hfill}
}
\hfill
\vtop{\leftskip 0pt\parindent 0pt\hsize=85pt
\ligne{\hfill cell 1(5)\hfill}
\ligne{\schrule {Y} {WYWGMYY} {Y} {\ftt{357}} \hfill}
\ligne{\schrule {Y} {WBWGMYY} {B} {\ftt{255}} \hfill}
\ligne{\schrule {B} {WGMYYWR} {R} {\ftt{256}} \hfill}
\ligne{\schrule {R} {WYWGMYB} {Y} {\ftt{262}} \hfill}
\ligne{\schrule {Y} {WYWGMYR} {Y} {\ftt{379}} \hfill}
}
\hfill}
\vskip 5pt
\ligne{\hfill decrementing, $n>1$\hfill}
\ligne{\hfill
\vtop{\leftskip 0pt\parindent 0pt\hsize=85pt
\ligne{\hfill cell 0\hfill}
\ligne{\schrule {W} {WYYWYYY} {W} {\ftt{274}} \hfill}
\ligne{\schrule {W} {WYYWGYY} {W} {\ftt{503}} \hfill}
\ligne{\schrule {W} {WYYWRRY} {W} {\ftt{375}} \hfill}
\ligne{\schrule {W} {WYYWYRG} {W} {\ftt{504}} \hfill}
\ligne{\schrule {W} {WYYWYYR} {W} {\ftt{273}} \hfill}
}
\hfill
\vtop{\leftskip 0pt\parindent 0pt\hsize=85pt
\ligne{\hfill cell 1(5)\hfill}
\ligne{\schrule {Y} {WYWGMYY} {Y} {\ftt{357}} \hfill}
\ligne{\schrule {Y} {WGWGMYY} {R} {\ftt{455}} \hfill}
\ligne{\schrule {R} {WRWGMYY} {R} {\ftt{453}} \hfill}
\ligne{\schrule {R} {WYWGMYG} {Y} {\ftt{465}} \hfill}
\ligne{\schrule {Y} {WYWGMYR} {Y} {\ftt{379}} \hfill}
}
\hfill}
}
\vskip 10pt
Table~\ref{t_debreg} can be checked on Figure~\ref{f_debreg}.

\vtop{
\ligne{\hfill
\includegraphics[scale=0.5]{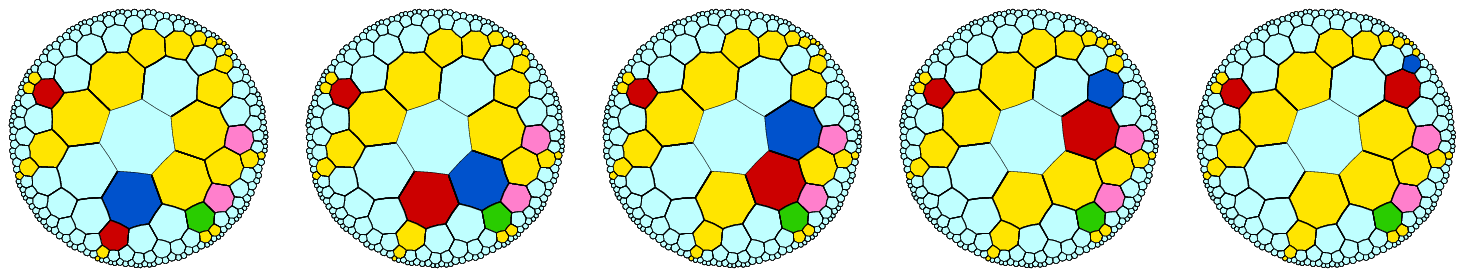}
\hfill}
\ligne{\hfill
\includegraphics[scale=0.5]{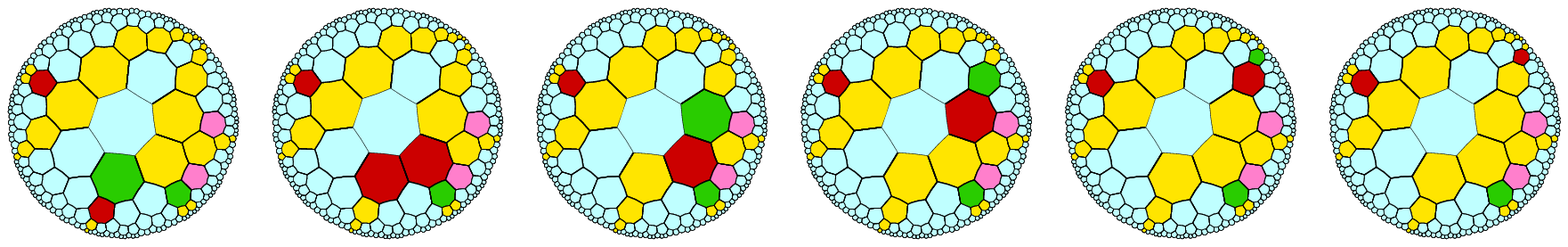}
\hfill}
\begin{fig}\label{f_m_debreg}
\leurre
Top, a \BB-locomotive going to increment $\mathcal R$ in the case when $n>0$.
Bottom, a \GG-locomotive going to decrement $\mathcal R$ in the case when $n>1$.
\end{fig}
}

   The case when $n>0$ is displayed by Table~\ref{t_op_gn}. The rules deal with an incrementing
operation and then with a decrementing one, that latter one for $n>1$.

As far as incrementing or decrementing the register cannot be performed by the simple action of the
front of the locomotive, we can avoid to introduce two new states. The special mark \MM{} for the 
end of the register and the two fronts for locomotive are enough to entail a different working
allowing to get the transformation suggested by Figure~\ref{f_reg_op}. The state \MM{} and the
\GG-front are attached to a decrementing operation while the state \MM{} when the \BB-front is 
present is attached to the incrementing one.

When \MM{} and \BB{} are close from each other, the \MM-cell becomes \BB-, which triggers the hat 
and neighbouring cells to become \BB- or \MM. The occurrence of the patterns \BB\BB{} and \BB\MM{}
in the neighbourhood of a \WW-cell allows that one to become \GG- and later to turn to \YY- in order
to build the new hat, see Figure~\ref{f_incg} and rules 292 and 304 and also rules 301 and 308.

When \MM{} and \GG{} are close from each other, it is possible to operate in another way: the
\MM-cell becomes \GG, which allows to destroy the hat, keeping the needed cells to get the new one.

In both cases of incrementing and decrementing, the operation must be performed in such a way that
a return locomotive with the appropriate colour goes back along the return track along the 
register.

In the Tables~\ref{t_op_gn}, \ref{t_z_op} and \ref{t_u_op} if $\xi_u$ is an indexed number of a
rule, the index~$u$ indicates the number of consecutive iterations of the rule. When there is no 
index, the rule applies once only.
\vskip 10pt
\vtop{ 
\vspace{-20pt}
\begin{tab}\label{t_op_gn}
\leurre
The rules for the situation when the \RR-cell is \reg{n} , with $n >0$ for an incrementing
instruction and with $n> 1$ for a decrementing one. 
\end{tab}
\ligne{\hfill incrementing the register \hfill}
\ligne{\hfill
\vtop{\leftskip 0pt\parindent 0pt\hsize=85pt
\ligne{\hfill cell 0\hfill}
\vskip 1pt
\ligne{\schrule {M} {WYYYYYY} {M} {\ftt{269}} \hfill}
\ligne{\schrule {M} {WBYYYYY} {B} {\ftt{277}} \hfill}
\ligne{\schrule {B} {WRMYYYY} {B} {\ftt{281}} \hfill}
\ligne{\schrule {B} {WYMMBBY} {W} {\ftt{290}} \hfill}
\ligne{\schrule {W} {WYYYMRB} {W} {\ftt{302}} \hfill}
\ligne{\schrule {W} {WYYYMYR} {W} {\ftt{311}} \hfill}
\ligne{\schrule {W} {WYYYMYY} {W} {\ftt{317}} \hfill}
}
\hfill
\vtop{\leftskip 0pt\parindent 0pt\hsize=85pt
\ligne{\hfill cell 1(6)\hfill}
\vskip 1pt
\ligne{\schrule {Y} {WWWWYMY} {Y} {\ftt{270$_2$}} \hfill}
\ligne{\schrule {Y} {WWWWYBM} {M} {\ftt{288}} \hfill}
\ligne{\schrule {M} {WWWWBBM} {Y} {\ftt{298}} \hfill}
\ligne{\schrule {Y} {WWWGMWY} {Y} {\ftt{306}} \hfill}
\ligne{\schrule {Y} {WWWYMWY} {Y} {\ftt{315$_2$}} \hfill}
}
\hfill}
\vskip 5pt
\ligne{\hfill decrementing the register \hfill}
\ligne{\hfill
\vtop{\leftskip 0pt\parindent 0pt\hsize=85pt
\ligne{\hfill cell 0\hfill}
\vskip 1pt
\ligne{\schrule {M} {WYYYYYY} {M} {\ftt{269}} \hfill}
\ligne{\schrule {M} {WGYYYYY} {G} {\ftt{387}} \hfill}
\ligne{\schrule {G} {WYWYGGY} {Y} {\ftt{394}} \hfill}
\ligne{\schrule {Y} {WWWWGMG} {Y} {\ftt{401}} \hfill}
\ligne{\schrule {Y} {WWWWRMR} {Y} {\ftt{406}} \hfill}
\ligne{\schrule {Y} {WWWWYMY} {Y} {\ftt{270}} \hfill}
}
\hfill
\vtop{\leftskip 0pt\parindent 0pt\hsize=85pt
\ligne{\hfill cell 1(4)\hfill}
\vskip 1pt
\ligne{\schrule {Y} {WWWYMWY} {Y} {\ftt{315}} \hfill}
\ligne{\schrule {Y} {WWWYMWG} {G} {\ftt{386}} \hfill}
\ligne{\schrule {G} {WWWYMWR} {R} {\ftt{389}} \hfill}
\ligne{\schrule {R} {WWWWGWY} {Y} {\ftt{87}} \hfill}
\ligne{\schrule {Y} {WWWWGWY} {G} {\ftt{75}} \hfill}
\ligne{\schrule {G} {WWWWYMY} {R} {\ftt{402}} \hfill}
\ligne{\schrule {R} {WWWWYMY} {Y} {\ftt{411}} \hfill}
\ligne{\schrule {Y} {WWWWYMY} {Y} {\ftt{270}} \hfill}
}
\hfill}
}
\vskip 10pt

\vtop{
\ligne{\hfill	
\includegraphics[scale=0.5]{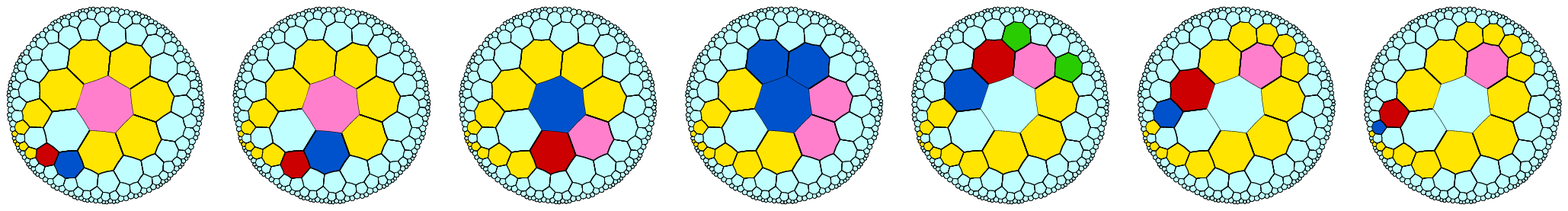}
\hfill}
\begin{fig}\label{f_incg}
\leurre
Execution of an incrementing instruction when $n>0$.
\end{fig}
}	

Figure~\ref{f_incg} illustrates the execution of an incrementing instruction on \reg{n} {}, with
$n>0$, so that the \RR-cell is transported to \reg{n+1} . Figure~\ref{f_decg} illustrates the case 
of a decrementing instruction, also when $n>0$. In Table~\ref{t_op_gn}, note that during the
decrementing operation, two rules, namely rule~87 and rule~75 are applied, rules that were
devised for the motion of a \GG-locomotive on the tracks. A look at Figures~\ref{f_decg} and
\ref{f_m_voies} shows us the similarity of configurations restricted to cell 1(4) of 
Figure~\ref{f_decg}. That observation reminds us that during an operation on a register all
rules of the table are available, in particular those which were defined for the tracks and for the
auxiliary mechanisms devised to simulate crossings and switches.

\vtop{
\ligne{\hfill	
\includegraphics[scale=0.5]{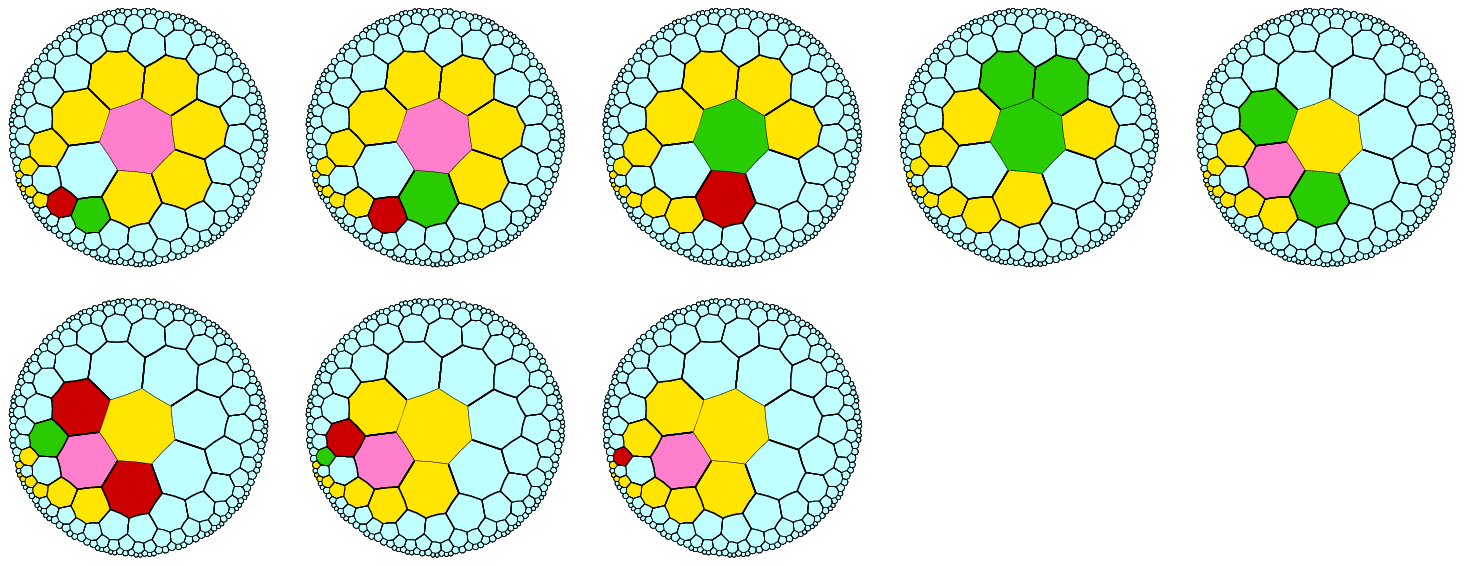}
\hfill}
\begin{fig}\label{f_decg}
\leurre
Execution of a decrementing instruction when $n>0$.
\end{fig}
}	

Table~\ref{t_z_op} deals with the case when the register is empty for an incrementing operation as
well as for a decrementing one.

Table~\ref{t_u_op} deals with the case of a decrementing operation in the case when $n=1$.

We can see in Table~\ref{t_op_gn} that the starting point of the operation is the same, so that
the same rule is applied.

\vskip 10pt
\vtop{
\begin{tab}\label{t_z_op}
\leurre
When the register is empty, so that ${\mathcal R}(0)$ is \MM.
\end{tab}
\ligne{\hfill increasing the register\hfill}
\ligne{\hfill
\vtop{\leftskip 0pt\parindent 0pt\hsize=85pt
\ligne{\hfill cell 0\hfill}
\ligne{\schrule {M} {WYYYYYY} {M} {\ftt{269}}\hfill}
\ligne{\schrule {M} {WBYYYYY} {B} {\ftt{277}}\hfill}
\ligne{\schrule {B} {WRMYYYY} {B} {\ftt{281}}\hfill}
\ligne{\schrule {B} {WYMMBBY} {W} {\ftt{290}}\hfill}
\ligne{\schrule {W} {WYYYMRB} {W} {\ftt{302}}\hfill}
\ligne{\schrule {W} {WYYYMYR} {W} {\ftt{311}}\hfill}
\ligne{\schrule {W} {WYYYMYY} {W} {\ftt{317}}\hfill}
}
\hfill
\vtop{\leftskip 0pt\parindent 0pt\hsize=85pt
\ligne{\hfill cell 1(4)\hfill}
\ligne{\schrule {Y} {WWYMWWB} {B} {\ftt{325}}\hfill}
\ligne{\schrule {B} {WWYMWWR} {R} {\ftt{335}}\hfill}
\ligne{\schrule {R} {WWYWWMB} {Y} {\ftt{339}}\hfill}
\ligne{\schrule {Y} {WWWYWWY} {Y} {\ftt{11$_3$}}\hfill}
}
\hfill
\vtop{\leftskip 0pt\parindent 0pt\hsize=85pt
\ligne{\hfill cell 1(7)\hfill}
\ligne{\schrule {Y} {WWWWYMY} {Y} {\ftt{270$_2$}}\hfill}
\ligne{\schrule {Y} {WWWWYBY} {B} {\ftt{282}}\hfill}
\ligne{\schrule {B} {WWWWBBM} {M} {\ftt{300}}\hfill}
\ligne{\schrule {M} {WWGRWYG} {M} {\ftt{307}}\hfill}
\ligne{\schrule {M} {WYYYYYY} {M} {\ftt{269}}\hfill}
}
\hfill}
\ligne{\hfill decreasing the register\hfill}
\ligne{\hfill
\vtop{\leftskip 0pt\parindent 0pt\hsize=85pt
\ligne{\hfill cell 0\hfill}
\ligne{\schrule {M} {WYYYYYY} {M} {\ftt{269}}\hfill}
\ligne{\schrule {M} {WGYYYYY} {G} {\ftt{387}}\hfill}
\ligne{\schrule {G} {WRGYYYY} {M} {\ftt{422}}\hfill}
\ligne{\schrule {M} {WYRYGGY} {M} {\ftt{432}}\hfill}
\ligne{\schrule {M} {WYYYYYY} {M} {\ftt{269}}\hfill}
}
\hfill
\vtop{\leftskip 0pt\parindent 0pt\hsize=85pt
\ligne{\hfill cell 1(5)\hfill}
\ligne{\schrule {Y} {WGMYYMY} {Y} {\ftt{326}}\hfill}
\ligne{\schrule {Y} {WGMYYMG} {G} {\ftt{420}}\hfill}
\ligne{\schrule {G} {WGMYYGR} {R} {\ftt{426}}\hfill}
\ligne{\schrule {R} {WGMGYMY} {Y} {\ftt{436}}\hfill}
\ligne{\schrule {Y} {WGMRYMY} {Y} {\ftt{444}}\hfill}
\ligne{\schrule {Y} {WGMYYMY} {Y} {\ftt{326}}\hfill}
}
\hfill
\vtop{\leftskip 0pt\parindent 0pt\hsize=85pt
\ligne{\hfill cell 2(6)\hfill}
\ligne{\schrule {Y} {WWYMYYM} {Y} {\ftt{330$_2$}}\hfill}
\ligne{\schrule {Y} {WWYMYGM} {G} {\ftt{431}}\hfill}
\ligne{\schrule {G} {WWYMYRM} {R} {\ftt{440}}\hfill}
\ligne{\schrule {R} {WGMYYMY} {Y} {\ftt{447}}\hfill}
\ligne{\schrule {Y} {WWRMYYM} {Y} {\ftt{449}}\hfill}
\ligne{\schrule {Y} {WWYMYYM} {Y} {\ftt{330}}\hfill}
}
\hfill}
}

\vskip 10pt
As already noticed about Figure~\ref{f_reg_op}, operations on the register destroy the existing hat
and restore it around the new place of the \MM-cell. Assuming the \MM-cell to be initially at~0,
it is at 1(7) after incrementing and it is at 1(3) after decrementing. That operation is driven by
the \MM-cells combined with \BB- or \GG-cells in an incrementing or a decrementing operation 
respectively. Note that, after decrementing, the previous \MM-cell becomes a \YY-cell of the hat 
of the new \MM-cell.

Looking at the considered tables, it can be seen that a few rules occur in two tables: rule~269 
occurs in those tables as the idle neighbourhood forming the hat of the \MM-cell marking the value
contained in the register. That rules occurs when the \MM-cell is cell~0 and also, in the case 
when the \MM-cell is the cell 1(7) in Table~\ref{t_z_op} for an incrementing instruction and in
Table~\ref{t_u_op} for a decrementing instruction. Figure~\ref{f_reg_z} illustrates both operations
on the empty register.

\vtop{
\ligne{\hfill
\includegraphics[scale=0.5]{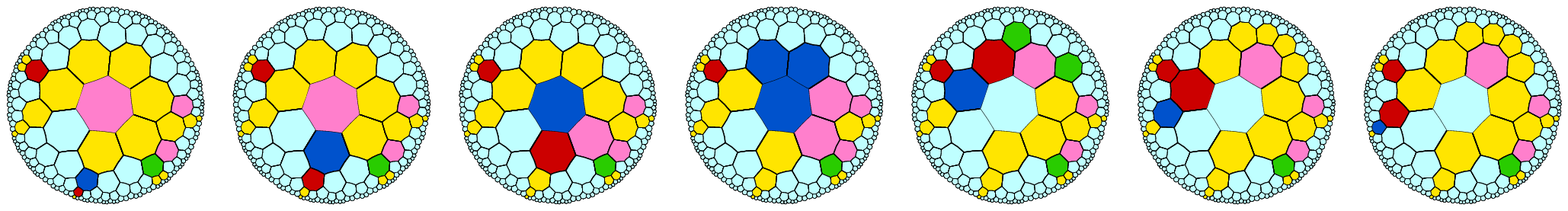}
\hfill}
\vspace{-10pt}
\ligne{\hfill
\includegraphics[scale=0.5]{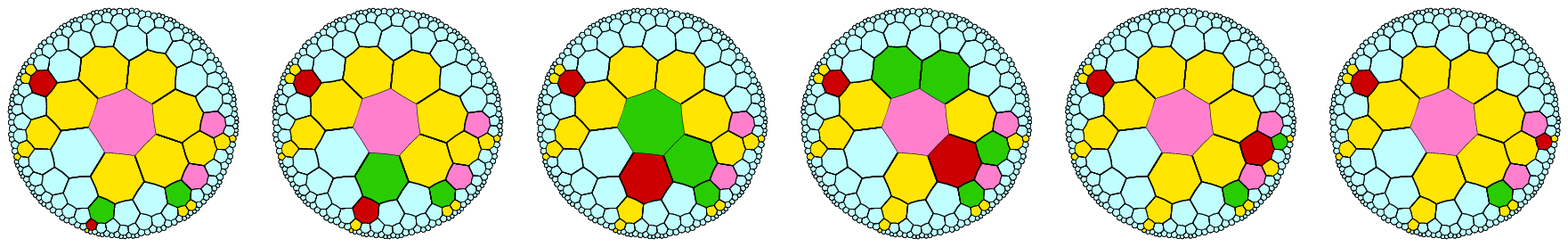}
\hfill}
\begin{fig}\label{f_reg_z}
\leurre
Empty register. Top, two rows: incrementing; bottom, one row: decrementing.
\end{fig}
}	

We can also note a change in the order of the states in a neighbourhood as far as a new state may
induce a change in the alphabetic order. As an example, the minimal form of rule~443, namely
\schrule {Y} {WWYWWYM} {Y} {} should be followed by \schrule {Y} {WWYWWYW} {Y} {} as far as the
state \MM{} becomes \WW. Now, the minimal form of that latter rule is \schrule {Y} {WWWYWWY} {Y} {}
which is rule~11, a rule for the tracks. That latter rule also draw our attention on a property of 
the tables: the rules were incrementally constructed: the rules displayed in Table~\ref{t_rules} 
for a given configuration are not exclusive from the rules established before for other 
configurations which can also be used in that context. 

\vtop{
\vspace{-10pt}
\begin{tab}\label{t_u_op}
\leurre
Decrementing the register when its content is $1$, so that ${\mathcal R}(2)$ is \MM.
\end{tab}
\ligne{\hfill
\vtop{\leftskip 0pt\parindent 0pt\hsize=85pt
\ligne{\hfill cell 0\hfill}
\ligne{\schrule {W} {WYYYMYY} {W} {\ftt{317}}\hfill}
\ligne{\schrule {W} {WGYYMYY} {W} {\ftt{453}}\hfill}
\ligne{\schrule {W} {WRRYMYY} {W} {\ftt{456}}\hfill}
\ligne{\schrule {W} {WYRGMYY} {W} {\ftt{388}}\hfill}
\ligne{\schrule {W} {WYYYGYY} {M} {\ftt{398}}\hfill}
\ligne{\schrule {M} {WYYYYYY} {M} {\ftt{269}}\hfill}
\ligne{\schrule {M} {WYYYMGY} {M} {\ftt{478}$_2$}\hfill}
\ligne{\schrule {M} {WYYYYRG} {M} {\ftt{494}}\hfill}
\ligne{\schrule {M} {WYYYYYR} {M} {\ftt{413}}\hfill}
\ligne{\schrule {W} {WYYYMYY} {W} {\ftt{317}}\hfill}
}
\hfill
\vtop{\leftskip 0pt\parindent 0pt\hsize=85pt
\ligne{\hfill cell 1(7)\hfill}
\ligne{\schrule {M} {WYYYYYY} {M} {\ftt{269$_3$}}\hfill}
\ligne{\schrule {M} {WGYYYYY} {G} {\ftt{387}}\hfill}
\ligne{\schrule {G} {WYWYYYY} {Y} {\ftt{473}}\hfill}
\ligne{\schrule {Y} {WYGGYMY} {M} {\ftt{476}}\hfill}
\ligne{\schrule {M} {WWGMYWY} {M} {\ftt{481}}\hfill}
\ligne{\schrule {M} {WWWYGMY} {Y} {\ftt{487}}\hfill}
\ligne{\schrule {Y} {WWWWMMY} {Y} {\ftt{493}}\hfill}
\ligne{\schrule {Y} {WWWWRMY} {Y} {\ftt{412}}\hfill}
\ligne{\schrule {Y} {WWWWYMY} {Y} {\ftt{270}$_2$}\hfill}
}
\hfill
\vtop{\leftskip 0pt\parindent 0pt\hsize=85pt
\ligne{\hfill cell 2(6)\hfill}
\ligne{\schrule {Y} {WWYMYYM} {Y} {\ftt{330$_2$}}\hfill}
\ligne{\schrule {Y} {WWYMYRM} {Y} {\ftt{446}}\hfill}
\ligne{\schrule {Y} {WWYMGRM} {Y} {\ftt{468}}\hfill}
\ligne{\schrule {Y} {WWYMYYM} {Y} {\ftt{330$_7$}}\hfill}
}
\hfill}
}
\vskip 10pt
Figure~\ref{f_decun} illustrates the application of the rules of Table~\ref{t_u_op}. Note that
the cell 2(6) behaves in another way that in Figure~\ref{f_reg_z}. When the \GG-locomotive is on 
the way to perform the decrementing instruction, the cell 2(6) remains \YY{} as far as the cell
1(5) changes from~\YY{} to \RR, as shown by rule~446. When the decrementing instruction cannot be 
performed, the cell 1(5) changes from~\YY{} to \GG, so that after rule~330, rule~431 applies which
allows the front of the \GG-locomotive to enter the cell 2(6). Rule 431 also shows us that the 
cell 1(6) remains \YY{} when the cell 2(6) becomes \GG. By contrast, rules~446 and 468 show us the
passage of the locomotive through the cell 1(6) when the \GG-locomotive is sent to decrement the
register in the case $n>0$.
\vskip 10pt
\vtop{
\ligne{\hfill
\includegraphics[scale=0.5]{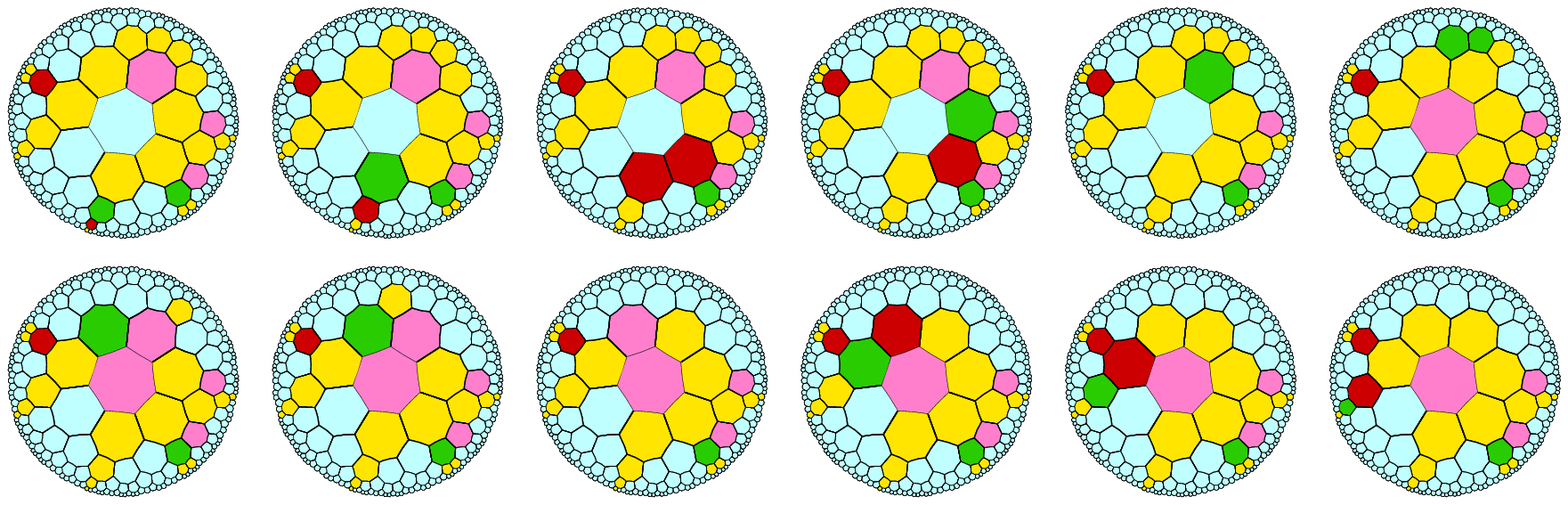}
\hfill}
\begin{fig}\label{f_decun}
\leurre
Decrementing a register whose content is $1$.
\end{fig}
}

Figure~\ref{f_decun} illustrates the transformation $n\rightarrow$~$n$$-$1 which transforms the
register to an empty one: the \MM-cell is moved from \reg{1} {} to \reg{0} {}.
\vskip 5pt
There are 510 rules in Table~\ref{t_rules} which were tested by a computer program. That 
program produced figures from the rules which allowed me to check their correction. As far as the 
stopping instruction is simply a track leading to a \GG-filter, the locomotive being then a \BB-one,
no additional rule is needed for that purpose.

Table~\ref{t_rules} displays the whole set of rules used by the automaton. We use the same 
convention as for the tables~\ref{t_move} up to~\ref{t_u_op}: each rules is displayed as
\schrule {c} {ABCDEFG} {n}, where $c$ is the current state, $n$ is the new state after the rule
The states taken by the neighbours of the cell are displayed in the alphabetic order in
{\ftt ABCDEFG}, which is not necessarily the order from neighbour~1 up to neighbour~7. 

\newcommand{\tttregle}[4]
{%
\hbox to 15pt{\hfill\ftt #1}
\hbox to 10pt{\hfill\ftt #2\hfill}
\hbox to 35pt{\hfill\ftt #3\hfill}
\hbox to 10pt{\hfill\ftt #4\hfill}
}

\begin{tab}\label{t_rules}
\leurre
Table of the rules used by the cellular automaton. 
\end{tab}

\ligne{\hfill
\vtop{\leftskip 0pt\parindent 0pt\hsize=82pt
\ligne{\hfill preliminary rules\hfill}
\ligne{\tttregle {  1} W {WWWWWWW} W \hfill}
\ligne{\tttregle {  2} W {WWWWWWY} W \hfill}
\ligne{\tttregle {  3} Y {WWWWWWW} Y \hfill}
\ligne{\tttregle {  4} W {WWWWWWB} W \hfill}
\ligne{\tttregle {  5} W {WWWWWWG} W \hfill}
\ligne{\tttregle {  6} W {WWWWWWR} W \hfill}
\ligne{\tttregle {  7} W {WWWWWBR} W \hfill}
\ligne{\tttregle {  8} W {WWWWWGR} W \hfill}
\ligne{\tttregle {  9} W {WWWWWRB} W \hfill}
\ligne{\tttregle { 10} W {WWWWWRG} W \hfill}
\ligne{\tttregle { 11} Y {WWWYWWY} Y \hfill}
\ligne{\tttregle { 12} W {WWWWWYY} W \hfill}
\ligne{\tttregle { 13} W {WWWWWYB} W \hfill}
\ligne{\tttregle { 14} W {WWWWWRY} W \hfill}
\ligne{\tttregle { 15} Y {WWWWWWY} Y \hfill}
}
\hfill
\vtop{\leftskip 0pt\parindent 0pt\hsize=82pt
\ligne{\hfill tracks\hfill}
\ligne{\hfill\ftt \BB-locomotive\hfill}
\ligne{\tttregle { 16} W {WWWWWBY} W \hfill}
\ligne{\tttregle { 17} B {WWWRWWY} R \hfill}
\ligne{\tttregle { 18} R {WWWYWWB} Y \hfill}
\ligne{\tttregle { 19} W {WWWWWYR} W \hfill}
\ligne{\tttregle { 20} Y {WWWWWWR} Y \hfill}
\ligne{\tttregle { 21} Y {WWWYWWR} Y \hfill}
\ligne{\tttregle { 22} Y {WWWWWWB} Y \hfill}
\ligne{\hfill\ftt \GG-locomotive\hfill}
\ligne{\tttregle { 23} W {WWWWWGY} W \hfill}
\ligne{\tttregle { 24} G {WWWRWWY} R \hfill}
\ligne{\tttregle { 25} R {WWWYWWG} Y \hfill}
\ligne{\tttregle { 26} W {WWWWWYG} W \hfill}
\ligne{\tttregle { 27} Y {WWWWWWG} Y \hfill}
}
\hfill
\vtop{\leftskip 0pt\parindent 0pt\hsize=82pt
\ligne{\hfill tracks on arc\hfill}
\ligne{\tttregle { 28} W {WWYYYYY} W \hfill}
\ligne{\hfill\ftt \BB-locomotive\hfill}
\ligne{\tttregle { 29} W {WWBYYYY} W \hfill}
\ligne{\tttregle { 30} W {WWRBYYY} W \hfill}
\ligne{\tttregle { 31} W {WWYRBYY} W \hfill}
\ligne{\tttregle { 32} W {WWYYRBY} W \hfill}
\ligne{\tttregle { 33} W {WWYYYRB} W \hfill}
\ligne{\tttregle { 34} W {WWYYYYR} W \hfill}
\ligne{\tttregle { 35} W {WWYYYYB} W \hfill}
\ligne{\tttregle { 36} Y {WWWWBWY} B \hfill}
\ligne{\tttregle { 37} W {WWYYYBR} W \hfill}
\ligne{\tttregle { 38} B {WWWWRWY} R \hfill}
\ligne{\tttregle { 39} W {WWYYBRY} W \hfill}
\ligne{\tttregle { 40} R {WWWWYWB} Y \hfill}
\ligne{\tttregle { 41} W {WWYBRYY} W \hfill}
\ligne{\tttregle { 42} Y {WWWWYWR} Y \hfill}
}
\hfill}

\ligne{\hfill
\vtop{\leftskip 0pt\parindent 0pt\hsize=82pt
\ligne{\hfill tracks on arc, cont.\hfill}
\ligne{\hfill\ftt \BB-locomotive, cont.\hfill}
\ligne{\tttregle { 43} W {WWWYYYB} W \hfill}
\ligne{\tttregle { 44} W {WWBRYYY} W \hfill}
\ligne{\tttregle { 45} W {WWRYYYY} W \hfill}
\ligne{\tttregle { 46} W {WWWYBRY} W \hfill}
\ligne{\tttregle { 47} W {WWWBRYY} W \hfill}
\ligne{\tttregle { 48} W {WWWRYYY} W \hfill}
\ligne{\tttregle { 49} W {WWWWYRB} W \hfill}
\ligne{\tttregle { 50} W {WWWBYYY} W \hfill}
\ligne{\tttregle { 51} W {WWWYYYY} W \hfill}
\ligne{\tttregle { 52} R {WWWWBWY} Y \hfill}
\ligne{\tttregle { 53} B {WWWYWWR} R \hfill}
\ligne{\tttregle { 54} Y {WWWWYWB} B \hfill}
\ligne{\tttregle { 55} Y {WWWWYWY} Y \hfill}
\ligne{\tttregle { 56} W {WWWWYYY} W \hfill}
\ligne{\tttregle { 57} W {WWWWYYR} W \hfill}
\ligne{\tttregle { 58} W {WWWRBYY} W \hfill}
\ligne{\tttregle { 59} Y {WWWWRWY} Y \hfill}
\ligne{\tttregle { 60} R {WWWBWWY} Y \hfill}
\ligne{\tttregle { 61} W {WWWYRBY} W \hfill}
\ligne{\tttregle { 62} Y {WWWRWWY} Y \hfill}
\ligne{\tttregle { 63} B {WWWWYWR} R \hfill}
\ligne{\tttregle { 64} Y {WWWYWWB} B \hfill}
\ligne{\tttregle { 65} W {WWWYYRB} W \hfill}
\ligne{\tttregle { 66} W {WWWYYYR} W \hfill}
\ligne{\tttregle { 67} W {WWWWBYY} W \hfill}
\ligne{\tttregle { 68} W {WWWWRBY} W \hfill}
\ligne{\hfill\ftt \GG-locomotive\hfill}
\ligne{\tttregle { 69} W {WWGYYYY} W \hfill}
\ligne{\tttregle { 70} W {WWRGYYY} W \hfill}
\ligne{\tttregle { 71} W {WWYRGYY} W \hfill}
\ligne{\tttregle { 72} W {WWYYRGY} W \hfill}
\ligne{\tttregle { 73} W {WWYYYRG} W \hfill}
\ligne{\tttregle { 74} W {WWYYYYG} W \hfill}
\ligne{\tttregle { 75} Y {WWWWGWY} G \hfill}
\ligne{\tttregle { 76} W {WWYYYGR} W \hfill}
\ligne{\tttregle { 77} G {WWWWRWY} R \hfill}
\ligne{\tttregle { 78} W {WWYYGRY} W \hfill}
\ligne{\tttregle { 79} R {WWWWYWG} Y \hfill}
\ligne{\tttregle { 80} W {WWYGRYY} W \hfill}
\ligne{\tttregle { 81} W {WWWGRYY} W \hfill}
\ligne{\tttregle { 82} W {WWWYYYG} W \hfill}
\ligne{\tttregle { 83} W {WWGRYYY} W \hfill}
\ligne{\tttregle { 84} W {WWWYGRY} W \hfill}
\ligne{\tttregle { 85} W {WWWWYRG} W \hfill}
}
\hfill
\vtop{\leftskip 0pt\parindent 0pt\hsize=82pt
\ligne{\tttregle { 86} W {WWWGYYY} W \hfill}
\ligne{\tttregle { 87} R {WWWWGWY} Y \hfill}
\ligne{\tttregle { 88} G {WWWYWWR} R \hfill}
\ligne{\tttregle { 89} Y {WWWWYWG} G \hfill}
\ligne{\tttregle { 90} W {WWWRGYY} W \hfill}
\ligne{\tttregle { 91} R {WWWGWWY} Y \hfill}
\ligne{\tttregle { 92} W {WWWYRGY} W \hfill}
\ligne{\tttregle { 93} G {WWWWYWR} R \hfill}
\ligne{\tttregle { 94} Y {WWWYWWG} G \hfill}
\ligne{\tttregle { 95} W {WWWYYRG} W \hfill}
\ligne{\tttregle { 96} W {WWWWGYY} W \hfill}
\ligne{\tttregle { 97} W {WWWWRGY} W \hfill}
\ligne{\hfill fork\hfill}
\ligne{\tttregle { 98} Y {WWYWYWY} Y \hfill}
\ligne{\hfill\ftt \BB-locomotive\hfill}
\ligne{\tttregle { 99} Y {WWBWYWY} B \hfill}
\ligne{\tttregle {100} W {WWWWYYB} W \hfill}
\ligne{\tttregle {101} B {WWRWYWY} R \hfill}
\ligne{\tttregle {102} W {WWWWYBY} W \hfill}
\ligne{\tttregle {103} W {WWWWYBR} W \hfill}
\ligne{\tttregle {104} R {WWYWBWB} Y \hfill}
\ligne{\tttregle {105} W {WWWWBRB} W \hfill}
\ligne{\tttregle {106} W {WWWWBRY} W \hfill}
\ligne{\tttregle {107} Y {WWYWRWR} Y \hfill}
\ligne{\tttregle {108} W {WWWWRYR} W \hfill}
\ligne{\tttregle {109} W {WWWWRYY} W \hfill}
\ligne{\tttregle {110} R {WWWYWWY} Y \hfill}
\ligne{\hfill\ftt \GG-locomotive\hfill}
\ligne{\tttregle {111} Y {WWGWYWY} G \hfill}
\ligne{\tttregle {112} W {WWWWYYG} W \hfill}
\ligne{\tttregle {113} G {WWRWYWY} R \hfill}
\ligne{\tttregle {114} W {WWWWYGY} W \hfill}
\ligne{\tttregle {115} W {WWWWYGR} W \hfill}
\ligne{\tttregle {116} R {WWYWGWG} Y \hfill}
\ligne{\tttregle {117} W {WWWWGRG} W \hfill}
\ligne{\tttregle {118} W {WWWWGRY} W \hfill}
\ligne{\hfill fixed switch\hfill}
\ligne{\tttregle {119} Y {WWYWYRY} Y \hfill}
\ligne{\tttregle {120} R {WWWWYYY} R \hfill}
\ligne{\hfill\ftt from left\hfill}
\ligne{\hfill\ftt \BB-locomotive\hfill}
\ligne{\tttregle {121} Y {WWYRWWB} B \hfill}
\ligne{\tttregle {122} Y {WWYWWRY} Y \hfill}
\ligne{\tttregle {123} Y {WWYWYRB} B \hfill}
\ligne{\tttregle {124} R {WWWWBYY} R \hfill}
\ligne{\tttregle {125} B {WWYRWWR} R \hfill}
}
\hfill
\vtop{\leftskip 0pt\parindent 0pt\hsize=82pt
\ligne{\tttregle {126} B {WWYWYRR} R \hfill}
\ligne{\tttregle {127} R {WWWWRBY} R \hfill}
\ligne{\tttregle {128} R {WWYWWBR} Y \hfill}
\ligne{\tttregle {129} W {WWWWWRR} W \hfill}
\ligne{\tttregle {130} Y {WWYWWRB} Y \hfill}
\ligne{\tttregle {131} R {WWBWYRY} Y \hfill}
\ligne{\tttregle {132} R {WWWWYRY} R \hfill}
\ligne{\tttregle {133} Y {WWYWWRR} Y \hfill}
\ligne{\tttregle {134} Y {WWRWYRY} Y \hfill}
\ligne{\tttregle {135} Y {WWYWWYR} Y \hfill}
\ligne{\hfill\ftt \GG-locomotive\hfill}
\ligne{\tttregle {136} Y {WWYRWWG} G \hfill}
\ligne{\tttregle {137} Y {WWYWYRG} G \hfill}
\ligne{\tttregle {138} R {WWWWGYY} R \hfill}
\ligne{\tttregle {139} G {WWYRWWR} R \hfill}
\ligne{\tttregle {140} G {WWYWYRR} R \hfill}
\ligne{\tttregle {141} R {WWWWRGY} R \hfill}
\ligne{\tttregle {142} R {WWYWWGR} Y \hfill}
\ligne{\tttregle {143} Y {WWYWWRG} Y \hfill}
\ligne{\tttregle {144} R {WWGWYRY} Y \hfill}
\ligne{\hfill\ftt from right\hfill}
\ligne{\hfill\ftt \BB-locomotive\hfill}
\ligne{\tttregle {145} Y {WWBWWRY} B \hfill}
\ligne{\tttregle {146} Y {WWYWBRY} B \hfill}
\ligne{\tttregle {147} R {WWWWYYB} R \hfill}
\ligne{\tttregle {148} B {WWRWWRY} R \hfill}
\ligne{\tttregle {149} B {WWYWRRY} R \hfill}
\ligne{\tttregle {150} R {WWWWYBR} R \hfill}
\ligne{\tttregle {151} Y {WWYWWBR} Y \hfill}
\ligne{\tttregle {152} Y {WWWBWWY} B \hfill}
\ligne{\tttregle {153} R {WWYWWRB} Y \hfill}
\ligne{\hfill\ftt \GG-locomotive\hfill}
\ligne{\tttregle {154} Y {WWGWWRY} G \hfill}
\ligne{\tttregle {155} Y {WWYWGRY} G \hfill}
\ligne{\tttregle {156} R {WWWWYYG} R \hfill}
\ligne{\tttregle {157} G {WWRWWRY} R \hfill}
\ligne{\tttregle {158} G {WWYWRRY} R \hfill}
\ligne{\tttregle {159} R {WWWWYGR} R \hfill}
\ligne{\tttregle {160} Y {WWYWWGR} Y \hfill}
\ligne{\tttregle {161} Y {WWWGWWY} G \hfill}
\ligne{\tttregle {162} R {WWYWWRG} Y \hfill}
\ligne{\hfill converters\hfill}
\ligne{\hfill\ftt from \BB-locom. to \GG-one\hfill}
\ligne{\tttregle {163} Y {WWWYGWY} Y \hfill}
\ligne{\tttregle {164} G {WWYYWYY} G \hfill}
\ligne{\tttregle {165} Y {WWWWWYG} Y \hfill}
}
\hfill}

\ligne{\hfill
\vtop{\leftskip 0pt\parindent 0pt\hsize=82pt
\ligne{\tttregle {166} Y {WWWWWGY} Y \hfill}
\ligne{\tttregle {167} Y {WWBWWGY} G \hfill}
\ligne{\tttregle {168} Y {WWWGGWY} G \hfill}
\ligne{\tttregle {169} G {WWYGWYY} G \hfill}
\ligne{\tttregle {170} W {WWWWYGG} W \hfill}
\ligne{\tttregle {171} G {WWGYWWR} R \hfill}
\ligne{\tttregle {172} G {WWWRGWY} R \hfill}
\ligne{\tttregle {173} G {WWGRWYY} G \hfill}
\ligne{\tttregle {174} R {WWYWWGG} Y \hfill}
\ligne{\tttregle {175} R {WWWYGWG} Y \hfill}
\ligne{\tttregle {176} G {WWRYWYY} G \hfill}
\ligne{\tttregle {177} Y {WWWYGWR} Y \hfill}
\ligne{\tttregle {178} W {WWWWRYG} W \hfill}
\ligne{\tttregle {179} Y {WWYWWGY} Y \hfill}
\ligne{\hfill\ftt from \GG-locom. to \BB-one\hfill}
\ligne{\tttregle {180} Y {WWWYBWY} Y \hfill}
\ligne{\tttregle {181} B {WWYYWYY} B \hfill}
\ligne{\tttregle {182} Y {WWWWWYB} Y \hfill}
\ligne{\tttregle {183} Y {WWWWWBY} Y \hfill}
\ligne{\tttregle {184} Y {WWBYWWG} B \hfill}
\ligne{\tttregle {185} Y {WWWBBWY} B \hfill}
\ligne{\tttregle {186} B {WWYBWYY} B \hfill}
\ligne{\tttregle {187} W {WWWWYBB} W \hfill}
\ligne{\tttregle {188} B {WWBYWWR} R \hfill}
\ligne{\tttregle {189} B {WWWRBWY} R \hfill}
\ligne{\tttregle {190} B {WWBRWYY} B \hfill}
\ligne{\tttregle {191} R {WWYWWBB} Y \hfill}
\ligne{\tttregle {192} R {WWWYBWB} Y \hfill}
\ligne{\tttregle {193} B {WWRYWYY} B \hfill}
\ligne{\tttregle {194} Y {WWWYBWR} Y \hfill}
\ligne{\tttregle {195} W {WWWWRYB} W \hfill}
\ligne{\tttregle {196} Y {WWYWWBY} Y \hfill}
\ligne{\hfill \BB-filter\hfill}
\ligne{\hfill\ftt \BB-locomotive, OK\hfill}
\ligne{\tttregle {197} B {WYYWRYR} B \hfill}
\ligne{\tttregle {198} Y {WWRBRWY} Y \hfill}
\ligne{\tttregle {199} R {WWWWWBY} R \hfill}
\ligne{\tttregle {200} W {WWWYYBR} W \hfill}
\ligne{\tttregle {201} Y {WWWBWBY} B \hfill}
\ligne{\tttregle {202} W {WWWRBYB} W \hfill}
\ligne{\tttregle {203} R {WWWWWYB} R \hfill}
\ligne{\tttregle {204} B {WYBWRYR} B \hfill}
\ligne{\tttregle {205} B {WWWRWBY} R \hfill}
\ligne{\tttregle {206} W {WWWRBBR} W \hfill}
}
\hfill
\vtop{\leftskip 0pt\parindent 0pt\hsize=82pt
\ligne{\tttregle {207} B {WBRWRYR} B \hfill}
\ligne{\tttregle {208} W {WWWYBBR} W \hfill}
\ligne{\tttregle {209} R {WWWYWBB} Y \hfill}
\ligne{\tttregle {210} W {WWWRBRY} W \hfill}
\ligne{\tttregle {211} B {WRYWRYR} B \hfill}
\ligne{\tttregle {212} W {WWWBRBR} W \hfill}
\ligne{\tttregle {213} Y {WWWYWBR} Y \hfill}
\ligne{\tttregle {214} W {WWWRYBR} W \hfill}
\ligne{\tttregle {215} Y {WWWYWBY} Y \hfill}
\ligne{\hfill\ftt \GG-locomotive, no\hfill}
\ligne{\tttregle {216} Y {WWWGWBY} Y \hfill}
\ligne{\tttregle {217} W {WWWRBYG} W \hfill}
\ligne{\tttregle {218} Y {WWWRWBY} Y \hfill}
\ligne{\tttregle {219} W {WWWRBYR} W \hfill}
\ligne{\hfill \GG-filter\hfill}
\ligne{\hfill\ftt \GG-locomotive, OK\hfill}
\ligne{\tttregle {220} G {WYYWRYR} G \hfill}
\ligne{\tttregle {221} Y {WWRGRWY} Y \hfill}
\ligne{\tttregle {222} R {WWWWWGY} R \hfill}
\ligne{\tttregle {223} W {WWWYYGR} W \hfill}
\ligne{\tttregle {224} Y {WWWGWGY} G \hfill}
\ligne{\tttregle {225} W {WWWRGYG} W \hfill}
\ligne{\tttregle {226} R {WWWWWYG} R \hfill}
\ligne{\tttregle {227} G {WYGWRYR} G \hfill}
\ligne{\tttregle {228} G {WWWRWGY} R \hfill}
\ligne{\tttregle {229} W {WWWRGGR} W \hfill}
\ligne{\tttregle {230} G {WGRWRYR} G \hfill}
\ligne{\tttregle {231} W {WWWYGGR} W \hfill}
\ligne{\tttregle {232} R {WWWYWGG} Y \hfill}
\ligne{\tttregle {233} W {WWWRGRY} W \hfill}
\ligne{\tttregle {234} G {WRYWRYR} G \hfill}
\ligne{\tttregle {235} W {WWWGRGR} W \hfill}
\ligne{\tttregle {236} Y {WWWYWGR} Y \hfill}
\ligne{\tttregle {237} W {WWWRYGR} W \hfill}
\ligne{\tttregle {238} Y {WWWYWGY} Y \hfill}
\ligne{\hfill\ftt \BB-locomotive, no\hfill}
\ligne{\tttregle {239} Y {WWWBWGY} Y \hfill}
\ligne{\tttregle {240} W {WWWRGYB} W \hfill}
\ligne{\tttregle {241} Y {WWWRWGY} Y \hfill}
\ligne{\tttregle {242} W {WWWRGYR} W \hfill}
}
\hfill
\vtop{\leftskip 0pt\parindent 0pt\hsize=82pt
\ligne{\hfill changing filters\hfill}
\ligne{\hfill\ftt from \BB- to \GG-\hfill}
\ligne{\tttregle {243} Y {WWRBRWB} G \hfill}
\ligne{\tttregle {244} W {WWWWBYR} W \hfill}
\ligne{\tttregle {245} B {WYYWRGR} G \hfill}
\ligne{\tttregle {246} G {WWRBRWR} Y \hfill}
\ligne{\tttregle {247} W {WWWWRGR} W \hfill}
\ligne{\tttregle {248} R {WWWWWBG} R \hfill}
\ligne{\tttregle {249} R {WWWWWGB} R \hfill}
\ligne{\hfill\ftt from \GG- to \BB-\hfill}
\ligne{\tttregle {250} Y {WWRGRWB} B \hfill}
\ligne{\tttregle {251} G {WYYWRBR} B \hfill}
\ligne{\tttregle {252} B {WWRGRWR} Y \hfill}
\ligne{\tttregle {253} W {WWWWRBR} W \hfill}
\ligne{\hfill registers\hfill}
\ligne{\hfill\ftt incrementing\hfill}
\ligne{\hfill\ftt across $\mathcal R$(0) to $\mathcal R$$(n)$\hfill}
\ligne{\tttregle {254} W {WWWBWYY} W \hfill}
\ligne{\tttregle {255} Y {WBWGMYY} B \hfill}
\ligne{\tttregle {256} B {WGMYYWR} R \hfill}
\ligne{\tttregle {257} W {WWWWGBR} W \hfill}
\ligne{\tttregle {258} G {WWYYWMB} G \hfill}
\ligne{\tttregle {259} M {WWWWYBG} M \hfill}
\ligne{\tttregle {260} Y {WWBYMWY} B \hfill}
\ligne{\tttregle {261} Y {WWYMYBM} Y \hfill}
\ligne{\tttregle {262} R {WYWGMYB} Y \hfill}
\ligne{\tttregle {263} B {WWRYMWY} R \hfill}
\ligne{\tttregle {264} Y {WWYMBRM} Y \hfill}
\ligne{\tttregle {265} M {WWWWBYY} M \hfill}
\ligne{\tttregle {266} W {WWWWYBM} W \hfill}
\ligne{\tttregle {267} R {WWYYMWB} Y \hfill}
\ligne{\tttregle {268} W {WWWWBRM} W \hfill}
\ligne{\hfill\ftt $n \rightarrow n$+1,\hfill}
\ligne{\hfill\ftt $n>0$\hfill}
\ligne{\tttregle {269} M {WYYYYYY} M \hfill}
\ligne{\tttregle {270} Y {WWWWYMY} Y \hfill}
\ligne{\tttregle {271} Y {WWWYWMY} Y \hfill}
\ligne{\tttregle {272} W {WRBYMYY} W \hfill}
\ligne{\tttregle {273} W {WYYWYYR} W \hfill}
\ligne{\tttregle {274} W {WYYWYYY} W \hfill}
\ligne{\tttregle {275} W {WWWWYWY} W \hfill}
\ligne{\tttregle {276} Y {WWWYMWB} B \hfill}
\ligne{\tttregle {277} M {WBYYYYY} B \hfill}
\ligne{\tttregle {278} W {WYRBMYY} W \hfill}
\ligne{\tttregle {279} B {WWWYMWR} R \hfill}
}
\hfill}

\ligne{\hfill
\vtop{\leftskip 0pt\parindent 0pt\hsize=82pt
\ligne{\tttregle {280} Y {WWWWYMB} M \hfill}
\ligne{\tttregle {281} B {WRMYYYY} B \hfill}
\ligne{\tttregle {282} Y {WWWWYBY} B \hfill}
\ligne{\tttregle {283} W {WYYRBYY} W \hfill}
\ligne{\tttregle {284} R {WWWMBWY} Y \hfill}
\ligne{\tttregle {285} M {WWWWYBR} M \hfill}
\ligne{\tttregle {286} W {WWWWWMR} W \hfill}
\ligne{\tttregle {287} W {WWWWWWM} W \hfill}
\ligne{\tttregle {288} Y {WWWWYBM} M \hfill}
\ligne{\tttregle {289} W {WWWWWYM} W \hfill}
\ligne{\tttregle {290} B {WYMMBBY} W \hfill}
\ligne{\tttregle {291} B {WWWWYBB} R \hfill}
\ligne{\tttregle {292} W {WWWWWBB} G \hfill}
\ligne{\tttregle {293} Y {WWWYWBB} B \hfill}
\ligne{\tttregle {294} W {WYYYBYY} W \hfill}
\ligne{\tttregle {295} Y {WWWMBWY} Y \hfill}
\ligne{\tttregle {296} M {WWWWMBY} Y \hfill}
\ligne{\tttregle {297} W {WWWWWMY} W \hfill}
\ligne{\tttregle {298} M {WWWWBBM} Y \hfill}
\ligne{\tttregle {299} W {WWWWWMM} W \hfill}
\ligne{\tttregle {300} B {WWWWBBM} M \hfill}
\ligne{\tttregle {301} W {WWWWWBM} G \hfill}
\ligne{\tttregle {302} W {WYYYMRB} W \hfill}
\ligne{\tttregle {303} R {WWWBWMG} Y \hfill}
\ligne{\tttregle {304} G {WWWWWRM} Y \hfill}
\ligne{\tttregle {305} W {WYYYWBY} W \hfill}
\ligne{\tttregle {306} Y {WWWGMWY} Y \hfill}
\ligne{\tttregle {307} M {WWGRWYG} M \hfill}
\ligne{\tttregle {308} G {WWWWWMY} Y \hfill}
\ligne{\tttregle {309} W {WWWWWMG} Y \hfill}
\ligne{\tttregle {310} W {WWWWWGM} Y \hfill}
\ligne{\tttregle {311} W {WYYYMYR} W \hfill}
\ligne{\tttregle {312} Y {WWWRWMY} Y \hfill}
\ligne{\tttregle {313} Y {WWWWWYM} Y \hfill}
\ligne{\tttregle {314} W {WYYYWRB} W \hfill}
\ligne{\tttregle {315} Y {WWWYMWY} Y \hfill}
\ligne{\tttregle {316} M {WWYYWYY} M \hfill}
\ligne{\tttregle {317} W {WYYYMYY} W \hfill}
\ligne{\tttregle {318} W {WYYYWYR} W \hfill}
\ligne{\tttregle {319} W {WYYYWRY} W \hfill}
\ligne{\hfill\ftt $0 \rightarrow 1$\hfill}
\ligne{\tttregle {320} Y {WYWMYWR} Y \hfill}
\ligne{\tttregle {321} R {WWYWWYY} R \hfill}
\ligne{\tttregle {322} Y {WWWWWYR} Y \hfill}
\ligne{\tttregle {323} Y {WWWWWRY} Y \hfill}
\ligne{\tttregle {324} W {WWWYMYY} W \hfill}
}
\hfill
\vtop{\leftskip 0pt\parindent 0pt\hsize=82pt
\ligne{\tttregle {325} Y {WWYMWWB} B \hfill}
\ligne{\tttregle {326} Y {WGMYYMY} Y \hfill}
\ligne{\tttregle {327} G {WWYYWMY} G \hfill}
\ligne{\tttregle {328} M {WWWWYYG} M \hfill}
\ligne{\tttregle {329} W {WWWWMGY} W \hfill}
\ligne{\tttregle {330} Y {WWYMYYM} Y \hfill}
\ligne{\tttregle {331} Y {WWYWYYM} Y \hfill}
\ligne{\tttregle {332} M {WWWWYYY} M \hfill}
\ligne{\tttregle {333} W {WWWWMYY} W \hfill}
\ligne{\tttregle {334} W {WWWBMYY} W \hfill}
\ligne{\tttregle {335} B {WWYMWWR} R \hfill}
\ligne{\tttregle {336} Y {WGMYYMB} M \hfill}
\ligne{\tttregle {337} W {WWWWGYB} W \hfill}
\ligne{\tttregle {338} Y {WYWBYWR} Y \hfill}
\ligne{\tttregle {339} R {WWYWWMB} Y \hfill}
\ligne{\tttregle {340} M {WGMYYBR} M \hfill}
\ligne{\tttregle {341} W {WWWWGMR} W \hfill}
\ligne{\tttregle {342} G {WWYYWMM} G \hfill}
\ligne{\tttregle {343} M {WWWWYMG} M \hfill}
\ligne{\tttregle {344} Y {WWYBMYM} M \hfill}
\ligne{\tttregle {345} Y {WWYMYMM} Y \hfill}
\ligne{\tttregle {346} Y {WYWBBWR} B \hfill}
\ligne{\tttregle {347} W {WWWYBYY} W \hfill}
\ligne{\tttregle {348} Y {WWYWWMB} Y \hfill}
\ligne{\tttregle {349} M {WGMYMBY} Y \hfill}
\ligne{\tttregle {350} W {WWWWGMY} W \hfill}
\ligne{\tttregle {351} M {WWBBMYM} Y \hfill}
\ligne{\tttregle {352} Y {WWYMMMM} Y \hfill}
\ligne{\tttregle {353} M {WWWWMYY} M \hfill}
\ligne{\tttregle {354} B {WWRWRWY} R \hfill}
\ligne{\tttregle {355} R {WWYYWWB} R \hfill}
\ligne{\tttregle {356} W {WWWYWBY} W \hfill}
\ligne{\tttregle {357} Y {WYWGMYY} Y \hfill}
\ligne{\tttregle {358} Y {WYYMWGM} Y \hfill}
\ligne{\tttregle {359} W {WWWWGYM} W \hfill}
\ligne{\tttregle {360} R {WWYWRWB} Y \hfill}
\ligne{\tttregle {361} W {WWWWRRY} W \hfill}
\ligne{\tttregle {362} R {WWYYWWR} R \hfill}
\ligne{\tttregle {363} W {WWWWBRR} W \hfill}
\ligne{\tttregle {364} W {WWWYWRB} W \hfill}
\ligne{\tttregle {365} Y {WYYMWYM} Y \hfill}
\ligne{\tttregle {366} W {WWWWYYM} W \hfill}
\ligne{\tttregle {367} W {WWWYWYR} W \hfill}
\ligne{\tttregle {368} Y {WWYWRWY} Y \hfill}
\ligne{\tttregle {369} W {WWWYWYY} W \hfill}
}
\hfill
\vtop{\leftskip 0pt\parindent 0pt\hsize=82pt
\ligne{\hfill\ftt back to program\hfill}
\ligne{\tttregle {370} Y {WWYYMWY} Y \hfill}
\ligne{\tttregle {371} W {WYYWRBY} W \hfill}
\ligne{\tttregle {372} W {WYYWBYY} W \hfill}
\ligne{\tttregle {373} W {WYYWYRB} W \hfill}
\ligne{\tttregle {374} Y {WWBWRWY} B \hfill}
\ligne{\hfill\ftt decrementing\hfill}
\ligne{\hfill\ftt across $\mathcal R$(0) to $\mathcal R$$(n)$\hfill}
\ligne{\tttregle {375} W {WYYWRRY} W \hfill}
\ligne{\tttregle {376} Y {WWRYMWY} G \hfill}
\ligne{\tttregle {377} W {WYYWYRY} W \hfill}
\ligne{\tttregle {378} G {WWRYMWY} R \hfill}
\ligne{\tttregle {379} Y {WYWGMYR} Y \hfill}
\ligne{\tttregle {380} R {WWYYMWG} Y \hfill}
\ligne{\tttregle {381} Y {WWYMRYM} Y \hfill}
\ligne{\tttregle {382} M {WWWWRYY} M \hfill}
\ligne{\tttregle {383} W {WWWWGRM} W \hfill}
\ligne{\tttregle {384} Y {WWYYMWR} Y \hfill}
\ligne{\hfill\ftt $n \rightarrow n$$-$1,\hfill}
\ligne{\hfill\ftt $n>1$\hfill}
\ligne{\tttregle {385} W {WRGYMYY} W \hfill}
\ligne{\tttregle {386} Y {WWWYMWG} G \hfill}
\ligne{\tttregle {387} M {WGYYYYY} G \hfill}
\ligne{\tttregle {388} W {WYRGMYY} W \hfill}
\ligne{\tttregle {389} G {WWWYMWR} R \hfill}
\ligne{\tttregle {390} Y {WWWWYMG} W \hfill}
\ligne{\tttregle {391} G {WYYYYWR} G \hfill}
\ligne{\tttregle {392} Y {WWWWYGY} G \hfill}
\ligne{\tttregle {393} W {WYYRGYY} W \hfill}
\ligne{\tttregle {394} G {WYWYGGY} Y \hfill}
\ligne{\tttregle {395} G {WWWWYGG} W \hfill}
\ligne{\tttregle {396} W {WWWWWGG} W \hfill}
\ligne{\tttregle {397} Y {WWWYWGG} G \hfill}
\ligne{\tttregle {398} W {WYYYGYY} M \hfill}
\ligne{\tttregle {399} Y {WWWWWGG} W \hfill}
\ligne{\tttregle {400} G {WWWWGGY} W \hfill}
\ligne{\tttregle {401} Y {WWWWGMG} Y \hfill}
\ligne{\tttregle {402} G {WWWWYMY} R \hfill}
\ligne{\tttregle {403} M {WYYGYGY} M \hfill}
\ligne{\tttregle {404} Y {WWWYWMG} G \hfill}
\ligne{\tttregle {405} Y {WWWWGMY} Y \hfill}
\ligne{\tttregle {406} Y {WWWWRMR} Y \hfill}
\ligne{\tttregle {407} R {WWWWGMY} Y \hfill}
\ligne{\tttregle {408} M {WYYRYRG} M \hfill}
\ligne{\tttregle {409} G {WWWYWMR} R \hfill}
\ligne{\tttregle {410} W {WYYYMGY} W \hfill}
}
\hfill}

\ligne{\hfill
\vtop{\leftskip 0pt\parindent 0pt\hsize=82pt
\ligne{\tttregle {411} R {WWWWYMY} Y \hfill}
\ligne{\tttregle {412} Y {WWWWRMY} Y \hfill}
\ligne{\tttregle {413} M {WYYYYYR} M \hfill}
\ligne{\tttregle {414} R {WWWGWMY} Y \hfill}
\ligne{\tttregle {415} W {WYYYMRG} W \hfill}
\ligne{\tttregle {416} W {WYYYWGY} W \hfill}
\ligne{\hfill\ftt decrementing\hfill}
\ligne{\hfill\ftt $n=0$, failure\hfill}
\ligne{\tttregle {417} Y {WWYMWWG} G \hfill}
\ligne{\tttregle {418} W {WWWGMYY} W \hfill}
\ligne{\tttregle {419} G {WWYMWWR} R \hfill}
\ligne{\tttregle {420} Y {WGMYYMG} G \hfill}
\ligne{\tttregle {421} W {WWWWGYG} W \hfill}
\ligne{\tttregle {422} G {WRGYYYY} M \hfill}
\ligne{\tttregle {423} Y {WYWGYWR} Y \hfill}
\ligne{\tttregle {424} R {WWYWYGG} Y \hfill}
\ligne{\tttregle {425} W {WWWWYRY} W \hfill}
\ligne{\tttregle {426} G {WGMYYGR} R \hfill}
\ligne{\tttregle {427} W {WWWWGGR} W \hfill}
\ligne{\tttregle {428} G {WWYYWMG} G \hfill}
\ligne{\tttregle {429} M {WWWWYGG} M \hfill}
\ligne{\tttregle {430} Y {WWYGGYM} Y \hfill}
\ligne{\tttregle {431} Y {WWYMYGM} G \hfill}
\ligne{\tttregle {432} M {WYRYGGY} M \hfill}
\ligne{\tttregle {433} G {WWWWYMG} Y \hfill}
\ligne{\tttregle {434} Y {WYWMGWR} Y \hfill}
\ligne{\tttregle {435} Y {WWYWWRM} Y \hfill}
\ligne{\tttregle {436} R {WGMGYMY} Y \hfill}
\ligne{\tttregle {437} G {WWYYWMR} G \hfill}
\ligne{\tttregle {438} M {WWWWGRG} M \hfill}
\ligne{\tttregle {439} Y {WWGMRGM} Y \hfill}
\ligne{\tttregle {440} G {WWYMYRM} R \hfill}
\ligne{\tttregle {441} M {WWWWYGY} M \hfill}
\ligne{\tttregle {442} G {WWWWGMY} Y \hfill}
\ligne{\tttregle {443} Y {WWYWWYM} Y \hfill}
\ligne{\tttregle {444} Y {WGMRYMY} Y \hfill}
\ligne{\tttregle {445} M {WWWYRYG} M \hfill}
\ligne{\tttregle {446} Y {WWYMYRM} Y \hfill}
\ligne{\tttregle {447} R {WGMYYMY} Y \hfill}
\ligne{\tttregle {448} Y {WWWWWRM} W \hfill}
\ligne{\tttregle {449} Y {WWRMYYM} Y \hfill}
\ligne{\tttregle {450} M {WWWWYYR} M \hfill}
\ligne{\tttregle {451} R {WWWYWMY} Y \hfill}
\ligne{\tttregle {452} W {WWWWMRY} W \hfill}
}
\hfill
\vtop{\leftskip 0pt\parindent 0pt\hsize=82pt
\ligne{\hfill\ftt $1 \rightarrow 0$\hfill}
\ligne{\tttregle {453} W {WGYYMYY} W \hfill}
\ligne{\tttregle {454} W {WWWGWYY} W \hfill}
\ligne{\tttregle {455} Y {WGWGMYY} R \hfill}
\ligne{\tttregle {456} W {WRRYMYY} W \hfill}
\ligne{\tttregle {457} W {WWWRWYY} W \hfill}
\ligne{\tttregle {458} R {WWWYWWR} Y \hfill}
\ligne{\tttregle {459} R {WGMYYWR} R \hfill}
\ligne{\tttregle {460} M {WWWWYRG} M \hfill}
\ligne{\tttregle {461} W {WWWWGRR} W \hfill}
\ligne{\tttregle {462} Y {WYMWRYM} G \hfill}
\ligne{\tttregle {463} Y {WWYWYRM} Y \hfill}
\ligne{\tttregle {464} W {WYYGMYY} W \hfill}
\ligne{\tttregle {465} R {WYWGMYG} Y \hfill}
\ligne{\tttregle {466} G {WYMWRYM} Y \hfill}
\ligne{\tttregle {467} Y {WWYWGRM} Y \hfill}
\ligne{\tttregle {468} Y {WWYMGRM} Y \hfill}
\ligne{\tttregle {469} M {WWWWGYY} M \hfill}
\ligne{\tttregle {470} W {WWWWYGM} W \hfill}
\ligne{\tttregle {471} Y {WYWGWYR} Y \hfill}
\ligne{\tttregle {472} Y {WWGWYYM} Y \hfill}
\ligne{\tttregle {473} G {WYWYYYY} Y \hfill}
\ligne{\tttregle {474} Y {WWWYMYG} G \hfill}
\ligne{\tttregle {475} G {WWWWYYG} W \hfill}
\ligne{\tttregle {476} Y {WYGGYMY} M \hfill}
\ligne{\tttregle {477} G {WWWWGYY} W \hfill}
\ligne{\tttregle {478} M {WYYYMGY} M \hfill}
\ligne{\tttregle {479} G {WWWWYMM} G \hfill}
\ligne{\tttregle {480} Y {WWMMYYM} Y \hfill}
\ligne{\tttregle {481} M {WWGMYWY} M \hfill}
\ligne{\tttregle {482} W {WWWWYMY} W \hfill}
\ligne{\tttregle {483} Y {WWWWWWM} W \hfill}
\ligne{\tttregle {484} G {WWWYMMY} M \hfill}
\ligne{\tttregle {485} Y {WWWWWGM} W \hfill}
\ligne{\tttregle {486} Y {WWYMMYY} Y \hfill}
}
\hfill
\vtop{\leftskip 0pt\parindent 0pt\hsize=82pt
\ligne{\tttregle {487} M {WWWYGMY} Y \hfill}
\ligne{\tttregle {488} Y {WWWWWMY} W \hfill}
\ligne{\tttregle {489} M {WYYYYMY} M \hfill}
\ligne{\tttregle {490} M {WWWWYMY} R \hfill}
\ligne{\tttregle {491} Y {WYWMMWR} G \hfill}
\ligne{\tttregle {492} W {WWWWRYM} W \hfill}
\ligne{\tttregle {493} Y {WWWWMMY} Y \hfill}
\ligne{\tttregle {494} M {WYYYYRG} M \hfill}
\ligne{\tttregle {495} G {WYWMRWR} R \hfill}
\ligne{\tttregle {496} R {WWYYWWG} R \hfill}
\ligne{\tttregle {497} W {WWWYMGY} W \hfill}
\ligne{\tttregle {498} R {WGWMYWR} Y \hfill}
\ligne{\tttregle {499} W {WWWYMRG} W \hfill}
\ligne{\tttregle {500} Y {WRWRWMY} Y \hfill}
\ligne{\tttregle {501} W {WWWYMYR} W \hfill}
\ligne{\hfill\ftt back to program\hfill}
\ligne{\tttregle {502} W {WYYWRGY} W \hfill}
\ligne{\tttregle {503} W {WYYWGYY} W \hfill}
\ligne{\tttregle {504} W {WYYWYRG} W \hfill}
\ligne{\tttregle {505} Y {WWGWRWY} G \hfill}
\ligne{\tttregle {506} W {WYYYWRG} W \hfill}
\ligne{\tttregle {507} G {WWRWRWY} R \hfill}
\ligne{\tttregle {508} W {WWWYWGY} W \hfill}
\ligne{\tttregle {509} R {WWYWRWG} Y \hfill}
\ligne{\tttregle {510} W {WWWYWRG} W \hfill}
}
\hfill}

\section{Conclusion}

We can see that we have 510 rules in the table, conservative, witness and motion rules. 
Note that the registers alone required 257 rules and 135 of those are required for
the special cases when the content of the register is 0 or 1. 
In Table~\ref{t_rules}, many rules are added to those of the previous tables which concerned a few
cells only while the additional rules of Table~\ref{t_rules} deal with all cells of the circuit.

We remain with the open question whether it is possible to reduce the number of
states. It most probably the case if we relax the condition of rotation invariance.
In that case, there is no need to distinguish between directions of the motion.

\end{document}